\documentclass[useAMS,usenatbib]{mn2e}
\usepackage{graphicx}
\begin{document}
\title[Encounter-induced brown dwarf formation]{A Numerical Study of
  Brown Dwarf Formation via Encounters of Protostellar Disks}
\author[S.~Shen et al.]{S.~Shen,$^1$\thanks{Email:
    shens@physics.mcmaster.ca} J.~Wadsley $^1$, T.~Hayfield $^1$ and
  N.Ellens $^1$ \\
 $^1$ Department of Physics and Astronomy, McMaster University, Main
 Street West, Hamilton L8S 4M1, Canada}
\maketitle 

\begin{abstract}

The formation of brown dwarfs (BDs) due to the fragmentation of
early-stage proto-stellar disks undergoing pairwise encounters was
investigated.  High resolution allowed the use of realistic, flared
initial disk models where both the vertical structure and the local
Jeans mass were resolved, preventing numerical enhancement or
suppression of fragmentation.  The results show that objects with
masses ranging from giant planets to low mass stars can form during
such encounters from disks that were stable before the encounter.  The
parameter space of initial spin-orbit orientations and the azimuthal
angles for each disk was explored with a series of 48 simulations.
For the types of interactions studied, an upper limit on the initial
Toomre Q value of $\sim 2$ was found for fragmentation to occur.  Depending
on the initial configuration, shocks, tidal-tail structures and mass
inflows were responsible for the condensation of disk gas into
self-gravitating objects.  In general, retrograde disks were more
likely to fragment.  It was also found that in fast encounters, where
the interaction timescale was significantly shorter than the disks'
dynamical timescales, the proto-stellar disks tended to be truncated
without forming objects.

The newly-formed objects had masses ranging from 0.9 to 127 $M_{J}$,
with the majority in the BD regime.  These often resided in star-BD
multiples and in some cases also formed hierarchical orbiting systems.
Most of them had large angular momenta and highly flattened, disk-like
shapes.  With the inclusion of the appropriate physics these could
reasonably be expected evolve into proto-BD disks with jets and
subsequent planet formation around the BD.  The objects had radii
ranging from 0.1 to 10 AU at the time of formation.  The disk gas was
assumed to be locally isothermal, appropriate for the short cooling
times in extended proto-stellar disks but not for condensed objects.
An additional case with explicit cooling that reduced to zero for
optically thick gas was simulated to test the extremes of cooling
effectiveness and it was still possible to form objects in this case.
Detailed radiative transfer (not studied here) is expected to lengthen
the internal evolution timescale for these objects so that they spend
considerable time as puffed-up, prolate ellipsoids, but not to alter
the basic result that proto-BD disks can form during proto-stellar
encounters.
 
\end{abstract}
\section{Introduction}

Observations suggest that brown dwarfs (BDs, 0.013 $M_{\sun} \la $ m
$\la 0.075 M_{\sun}$) are very abundant in both the field and star
clusters in our galaxy \citep{chabrier,martin,luhman}. 
The origin of brown dwarfs, however, is still under debate. It has been
suggested by several authors \citep[e.g.,][]{whitworth, bate02} that the final stellar mass 
may be closely related to the Jeans mass of the parent gas within the molecular 
cloud.  Given that the typical Jeans mass in a typical molecular cloud is of order
one solar mass, this is reasonable for larger, hydrogen-burning stars.
However, because BD masses are at least an order of
magnitude smaller, it is difficult for BDs to form by direct
collapse to pre-BD cores.  Several scenarios have been proposed to circumvent this problem.

One approach is to assume that the initial Jeans mass of the core is
still a good estimate of the final mass from a direct collapse, but
that denser or colder regions exist where the Jeans mass may approach 
the BD mass.  For example,
\citet{padoan} proposed that local density of in cores can be
enhanced by turbulent flows (``turbulent fragmentation''), while
\citet{elmegreen} suggested that certain types of molecular cooling
can greatly decrease core temperatures.  In either case, formation
of BDs follows the conventional star formation path. Hence, all
properties associated with proto-stars, such as accretion disks and
outflows, are expected.

An alternative picture, first proposed by \cite{reipurth}, suggests
that the because the Jeans mass continuously decreases during the
collapse, fragmentation can keep occurring repeatedly to produce
smaller and denser objects until they are too dense to have efficient
radiative cooling.  The mass at this ``opacity limit'' is expected to
be around $m_{\rm opacity} \sim M_{\rm Jupiter}$.  Thus the opacity limit mass,
rather than the initial core Jeans mass, would set the mass of the
smallest fragments.  In this scenario, it normally assumed that these
small masses form first.  These so-called embryos then accrete mass
and ultimately form both sub-stellar and stellar objects in a scenario
referred to as ``competitive accretion''.  The accretion of the
surrounding material onto embryos is assumed to be rapid so that
embryos remaining in dense parts of the molecular cloud would
ultimately become stellar-mass objects.  \citet{cluster_bate} explored
this scenario with a simulation of cluster formation from a 50
$M_{\sun}$ molecular cloud and found embryos could remain in the cloud
undergoing rapid accretion or get ejected during violent
interactions, resulting in a large number of low mass objects.  More
recent work indicates that a more realistic treatment of radiative
energy losses can strongly suppress the formation of sub-stellar
objects in such simulations \citep{pricebate}.  

The existence of extended disks around young stellar objects is
supported by both observations and theory \citep[e.g.][]{yorke}.  The
disks are initially large and flared at the class 0-I stage (extending
to several thousands AU \citep[see, e.g.,][]{mundy} and are observable
at mm or sub-mm wavelengths.  They evolve into smaller, thin
proto-planetary disks at the class III stage that are detectable with
infrared observations.  Disk lifetimes are inferred to be about 10 Myr.
Disks provide a natural environment for forming sub-stellar objects
because the Jeans mass in disks is much smaller than in cores.  It may
be possible to form brown dwarf mass companions through the
fragmentation of marginally stable ($Q \sim 1$), extended disks around single stars
\citep[e.g.,][]{stamatellos}.  Whether disks become cold and heavy
enough to spontaneously fragment depends on the disk mass build up
from the envelope and the processes that are driving accretion.
Self-regulated, stable ($Q > 1.5$), extended disks are the outcome for a significant
portion of the parameter space \citep{vorobyov}.

The approach that will be investigated in this paper is
encounter-triggered fragmentation of otherwise stable proto-stellar
disks.  Numerical studies by \citet{watkins, watkinsb} and \citet{lin}
have independently found that instabilities leading to bound
companions can be triggered during encounters between stable
proto-stellar disks.  Fragmentation was seen to occur within the disk
\citep{watkinsb}, or in filamentary structures induced by the tidal
force between the encountering disks \citep{lin}.  The resultant
objects reported by both groups include sub-stellar mass objects.

Since most stars form in clusters with a relatively small spread in
ages \citep{gauvin}, it is reasonably likely that a proto-star will
undergo an encounter during the time that the disks are large
\citep[e.g.][]{thies, watkins}.  Following the argument by
\citet{thies}, for a Plummer star cluster model with mass of $500
M_{\sun}$ and a half-mass radius, $R_{0.5} = 0.5 $ pc, the encounter
probability in $ 5 \times 10^{5}$ years (the approximate lifetime for
an early extended disk) with an impact parameter at pericenter of 500
AU is about 30 percent. If several sub-stellar objects form during
each encounter, as in \citet{watkins}, a substantial numbers of BDs
can be produced.  In this work, we focus on the outcomes of specific
encounters rather than estimates of their likelihood.

The prior simulation work failed to address several important issues.
Firstly, the resolution used in the work cited above did not ensure
that the Jeans mass was resolved in the original disks.  Thus it was
not clear whether the objects might have condensed due to numerical effects
\citep{bate_res, truelove}.  

Secondly, the range of circumstances in which gravitational
instabilities can be triggered in a stable disk by an encounter were
not explored.  The Toomre criterion for instability requires that that $ Q <
Q_{\rm crit}$ for thin disks to be unstable to axisymmetric perturbations,
where Q is the Toomre parameter, locally defined according to \citep{toomre},
\begin{equation}
Q \equiv \frac{c_{s} \kappa}{\pi G \Sigma},    \label{eq:toomreq} 
\end{equation}
where $c_{s}$ is the sound speed, $\kappa$ is the epicyclic frequency
(close to the angular speed $\Omega$ for a Keplerian disk) and
$\Sigma$ is the disk surface density.  Analytical studies by
\citet{toomre} and numerical studies of relatively thin proto-planetary
disks \citep{boss, mayer, pickett, gammie} 
have shown that $Q_{\rm crit} $ is in the range 1-1.5 for various modes of
fragmentation.  If proto-stellar disks are stable according to this
criterion then they tend to stay so indefinitely in the absence of
mechanisms to redistribute material.   A non-negligible
(vertical) thickness for the disks can be stabilizing, reducing the
critical Q value for axisymmetric perturbations from $1$ down to
$0.6-0.7$ \citep{kim}.  The evolution of the Toomre parameter during
dynamical interactions is a good indicator for the onset of
instabilities and fragmentation and this has been explored here.  It should
also be noted that the initial disk models assumed in prior work did
not resolve the disk vertical structure, which can destabilize the
disk (by increasing the effective critical Toomre Q) and artificially
enhance fragmentation.  These issues motivated our use of detailed,
vertically-resolved initial disk models.

Lastly, there have been few predictions regarding the properties of the
young BDs that could be produced during a proto-stellar disk
encounter.  Most observed young brown dwarfs show signatures of
surrounding disks analogous to the disks around proto-stars, and
some of these disks have been identified as accretors through spectral line
analysis \citep[e.g.][]{jay}.  The detection of BD accretion disks has
been used as an evidence to support the turbulent fragmentation
scenario
\citep[see, e.g.][]{padoan} against dynamical formation mechanisms,
especially after the aforementioned simulation by
\citet{cluster_bate} produced an practically no proto-BD
disks.  
Early encounter simulations, such as the ones by \citet{lin} and
\citet{watkins, watkinsb}, also reported no proto-BD disks.  However,
these negative results are quite likely to reflect insufficient resolution or the impact of
approximations such as replacing the fragments with a large
``sink'' particles \citep{cluster_bate}.  Studies with better resolved disks and
fragments were thus required to address this question.

The current study investigated encounter-induced BD formation using a
series of high resolution hydrodynamical simulations.  This allowed the
Jeans mass to be resolved until well after fragmentation initially took place and
the vertical structure of the initial disks was also well resolved.  The
paper is organized as follows: in section~\ref{methods} we describe
the numerical method, the construction of the initial disk models and
a study of resolution requirements.  In section~\ref{paramspace} we
present the encounter parameter space being explored, including 
Q values, relative orientations of disks and variations in
encounter velocity.  Section~\ref{results} describes several typical
simulations and discusses the conditions for encounter-induced fragmentation.  Section~\ref{objects}
presents analysis of the properties of the resultant objects, including shapes, angular
momenta and orbital evolution.  Section~\ref{discussion}
summarizes the results and includes discussion of the observational implications and
possible directions for future work.

\section{Methods}   \label{methods}

\subsection{Proto-Stellar Disk Model}

Observing the exact mass, density structure and kinematics of
proto-stellar objects during the earliest stages is difficult
\citep{white}.  Using a hydrodynamical calculation of proto-stellar
disk formation, \citet{yorke} found that early disks contain a large
fraction of the total mass of the system and collapse relatively
slowly.  Based on these results, we assumed that the proto-stellar
disks were quasi-hydrostatic in order to construct a self-consistent
disk density structure.

We chose to simulate systems in the solar mass range with a single
model for the disk-proto-star system where the disk mass was about 0.6
$M_{\sun}$, with a stellar mass of around 0.5 $M_{\sun}$.  
The disk extended to about 1000 AU. This
aspect of the set-up is similar to the disks simulated by
\citet{watkins, watkinsb}.  A power-law surface density profile
$\Sigma = \Sigma_{0}^{-p}$ was chosen with $p=1.5$ in 100-500 AU,
consistent with observations of early proto-stellar disks
\citep{andre}.  The surface density peaked at 100 AU with $\Sigma_{max}
\sim 20.0 \ g/cm^{3}$.  Within 100 AU
the gas was initially smoothly truncated.  We relaxed our initial
disks in isolation for about 10000 years, during which a
small amount of material occupied the truncated region.  Beyond 500 AU the
surface density falls rapidly but smoothly, and at 1000 AU $\Sigma
\sim 0.03\ g/cm^{3}$.  The entire profile after relaxation is similar
to the results from core collapse calculations by \citet{yorke}. Most of the disk
mass (0.5 $M_{\sun}$) is within 500 AU.

It was assumed that the central proto-star was the major heating source
for the disk, and that the disk re-radiates like a black-body, giving
a disk temperature profile of $T(r) \propto r^{-1/2}$ \citep{mayer,
pickett, chiang} for the majority of the simulations.  The profile
smoothly transitions to an external, ambient temperature of $\sim 40-80 $ K.
This profile is similar to the \citet{yorke} results.  With the radial
surface density and temperature profiles specified, the vertical
structure was iterated to a self-consistent hydrostatic equilibrium
taking into account both the gravity of the central star and the
self-gravity of the disk.  Accurate treatment of the gas self-gravity was
important given the large disk mass relative to the star.

\subsection{Radial Toomre Q Profile}

The choice of radial profile gave a nearly constant Q value
between 100-500 AU, with a minimum value ($Q_{\rm min}$) ranging
from 1.6 to 2.1 depending on the case studied.  The Q value rises in
the inner 100 AU and the disk is very stable there.  Test simulations
of single disk evolution with different values of $Q_{\rm min}$
were performed, and it was found that for our disk model values of $Q_{\rm min}$
below $Q_{\rm crit}= 0.8$ were required for spontaneous
fragmentation via spiral waves.  This critical value was lower than the
ones seen in thin disk models (e.g., $Q_{\rm crit} = 1.4$ in
\citet{mayer}), consistent with the results from \citet{kim} that
significant scale heights tend stabilize the disk.  To rule out
spontaneous fragmentation as a factor during the dynamical
interactions between disks, the $Q_{\rm min}$ values used were at
least twice as the $Q_{\rm crit}$ for an isolated disk, i.e. $Q_{\rm
  min} \geq 1.6$.

\subsection{Heating and Cooling Approximations}

A fixed temperature profile $T(r) \propto r^{-1/2}$ for most of the
disk was used in the
majority of our simulations, which is often referred to as a locally
isothermal equation of state (EOS).  Heating sources other than
the star, such as shocks, were not taken into account.  For the current
study, it was found that the isothermal EOS was a good approximation
for the initial disk and also acceptable even after disk was
heated by tidal structures and shocks (outside the clumps) because the
disks are extended and initially optically thin.  Given the temperature
of the gas, using the surface density profile described above and
adopting the opacities from \citet{dalessio}, the product of the disk
cooling time $t_{\rm cool}$ and angular rotation rate $\Omega$ was
calculated and $\Omega\ t_{\rm cool} = 0.3-0.7$ beyond 50 AU for the
initial disk, which is comfortably within the cooling requirement for
the disk to fragment $\Omega\ t_{\rm cool} < 12$ \citep{rice, johnson} before
the disk is dense enough for the isothermal assumption breaks down.
However, after the disk fragments and the resultant objects became
gravitationally bound, the final densities of these objects were typically orders
of magnitude larger than the initial values and the objects were thus
optically thick.  Thus our locally isothermal
approximation was initially good but later became an upper bound on
how efficient cooling should be.  To further examine the importance of
cooling in fragmentation, a case was also investigated with 
explicit cooling that became negligible in the optically thick
regime.  This second assumption is a lower bound for the cooling in a
real disk as it disallows even the slow leakage of radiation from
dense gas.  The results of this case are described in
section~\ref{cooling}.

\subsection{Simulations}

We used the TreeSPH code GASOLINE \citep{wadsley} to model the disks
during encounters.  Around 200,000 particles were used for each $0.6
M_{\sun}$ disk, with each particle representing $\sim$ 1 Earth mass.
This resolution was chosen to satisfy the condition that the local
Jeans mass should be resolved by at least $2 N_{neighbor}$
\citep{bate_res} to prevent numerical fragmentation, where
$N_{neighbor}$ is the number of neighbor particles in SPH (Smoothed
Particle Hydrodynamics) simulation. The number of particles ensured
that the Jeans mass was not-only well resolved in the initial
condition, but also well-resolved when density was enhanced during an
encounter (when the Jeans mass decreased), up to a density $\rho \sim
10^{-10}$ g cm$^{-3}$ (about $10^{3}$ times the initial maximum value
in the disk).  Resolving the Jeans mass within the fragments, however,
was not attempted due to high computing costs ($N > 10^{7}$ particles
usually required).  Therefore, our choice of particle numbers ensures
physically correct modeling of the initial fragmentation in the proto-stellar
disk, but cannot ensure that any secondary fragmentation of the clumps
themselves are physical.

The initial disk model also resolved the scale height at the mid-plane very
well at 100 AU (where the local smoothing length was about
1.9 AU and the scale height was about 6 AU).  Within 50 AU the scale
height of the disk was less well resolved and the orbital time was
shortest so that artificial viscous evolution could have been a
concern.  The artificial viscosity was moderated using the Balsara
switch which helps to suppress unwanted shear effects.  The measured viscous time
scale in the code was about $10^{8}$ to $10^{11}$ years, much longer
than the dynamical scale of the encounter, $t \sim 10^{5}$ years.
Hence, the effect due to the viscous evolution was negligible.  
Appendix \ref{resolution} presents as exploration of the effects of unresolved
Jeans mass and unresolved disk scale height.

The gravitational softening, $\epsilon$, for gas particles was 0.2 AU,
one tenth of the smallest particle spacing in the initial
condition. Thus dense regions up to 1000 times the initial densest
part were well resolved.  Since a larger $\epsilon$ usually
suppresses gravitation forces, the choice of gravitational softening
will tend to inhibit fragmentation once the particle separations are
comparable to or smaller than the softening.  In all tests, including
those done for this work and in other work to date, there have been no
indications that the Gasoline code suffers from artificial
fragmentation as a consequence of the gravitational softening being
smaller than the local SPH smoothing length.  The fixed softenings
ensure accurate gravitational dynamics at the cost of having somewhat
conservative time steps in lower density regions.

The central proto-star was represented by a single 0.5 solar mass sink
particle for which the softening was set to 1 AU.  The sink acted
within a 1 AU radius around the star to consume gas falling into the
innermost regions and thus avoid the small time steps required for gas
orbiting very close to the star.  However, no objects formed during
the simulation were replaced with sink particles, ensuring that any
disks surrounding the newly formed, sub-stellar object were modeled
directly.
  
\begin{figure*}
\includegraphics[width=\textwidth]{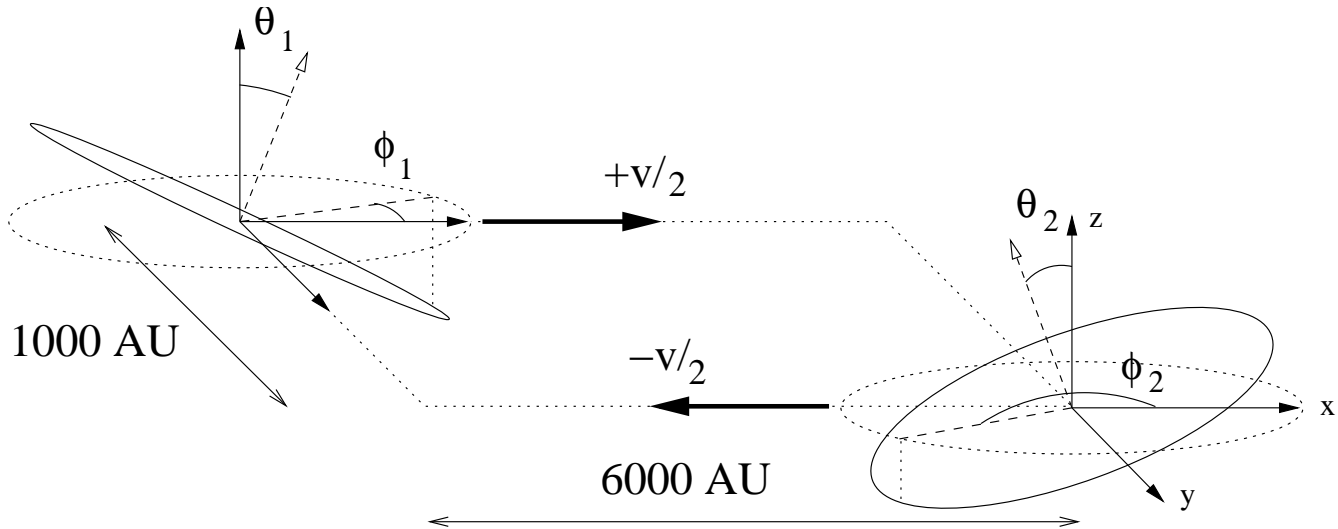}
\caption{A generic encounter configuration between two proto-stellar
  disks.  The centres of mass orbit in the x-y plane with an initial
  velocity solely in x-direction.  The angular
  momentum of an individual disk has a vector direction that differs from the
  orbital angular momentum (the z-axis) by an angle $\theta$.   The disk
  angular momentum vector projected onto the x-y plane makes an angle
  $\phi$ with the x-axis.  }
\label{diskconfig}
\end{figure*}   

\section{Parameter Space}
\label{paramspace}

\subsection{Locally Isothermal Simulations}
In each pairwise disk encounter, the two disks were
identical in size, mass and Toomre parameter with an
initial impact parameter of 1000 AU.  Due to gravitational focusing
during relatively slow encounters the minimum distance between two
stars could become substantially less than the initial impact
parameter. The initial disk separation was 6000 AU.  This system was
initially bound if the relative velocity of the encounter is less than
0.8 km/s and the final impact parameter is a function of the relative
velocity.  This study explored six dimensions in the overall parameter
space:

\begin{enumerate} 
\item The initial Toomre parameter profile characterized using the minimum value, $Q_{\rm min}$; 
\item The geometric configuration of the encounter, including the
  angles between the two disk planes and the angles between the disk planes and the plane of the orbit of the two disks
  (a total of four independent parameters); 
\item The relative velocity of the encounter. 
\end{enumerate}

Mass, disk size and impact parameter variations were not included to limit the study
to a manageable size.  Smaller impact parameters are possible but
considerably less likely.  These variations may be explored in future work.

A list of all the simulations is shown in Table \ref{list}. The first 3
cases (Q1, Q2, Q3) were carried out to study the effect of the
Toomre parameter on encounter-triggered fragmentation with different
initial minimum Q values of 1.6, 1.8 and 2.1, respectively.
A higher Q profile means a more stable initial
condition, for which more effort is required to trigger fragmentation.
We selected an encounter configuration where fragmentation was
likely (coplanar, retrograde disks) to first explore the threshold in initial
Toomre Q$_{\rm min}$, above which dynamical interactions would become insufficient to
trigger fragmentation.  This study helped to constrain the region
of interest within the parameter space.

Previous work by \citet{lin} indicated that fragmentation can take
place in tidal structures forming during encounters while
\citet{watkins,watkinsb} suggested disks would fragment.  The degree
of tidal distortion and angular momentum transport has been shown to
depend on the initial disk spin orientations (with respect to the
orbital angular momentum) \citep{toomre2}.  Various disk spin
orientations were used in our simulations including prograde cases
(where the angle between the angular momentum vector of the disks and
that of the orbit was less than 90 degrees), retrograde cases (where the
spin-orbit angle was greater than 90 degrees) and combinations of the
two.  

In most simulations the disks were not close to the same plane,
as would be expected for randomly oriented disks.  However, coplanar
encounters (where the spin-orbit angles are zero) were also simulated
as special cases to investigate the shocks generated in such encounters
and the possibility of shock fragmentation.  In every case the orbital plane
was the x-y plane so that the orbital angular momentum was in the direction of the z-axis.

In a completely generic encounter, as shown in
Figure~\ref{diskconfig}, the disks can be oriented randomly.  
The configuration of each
disk in the encounter can then be characterized by two angles: the angle
between the spin angular momentum of each disk and the positive z
direction, $\theta$, and the angle between the positive x direction and
the projection of disk angular momentum onto the x-y plane, $\phi$,
which also determines where that disk intersects the x-y plane.
The initial disk velocities are always along the x-axis and the initial impact
parameter is due to an offset in the y-direction.

Cases Conf1-48 explored the ($\theta_1$, $\phi_1$, $\theta_2$, $\phi_2$)
parameter space. For a single disk, the
angle parameters were selected such that the $\theta$
space (from 0 to 180 degrees) was divided evenly into three ranges and the $\phi$
space (from 0 to 360 degrees) into four for a total of 12 cases.
Naively there would have been 144 possible combinations of disks but
due to the symmetry of the encounters and the
equivalence when the two disks were exchanged in space, there were 42 unique collisions to simulate.
An additional six cases were simulated where at least one disk was exactly in
the orbital plane, to investigate the effects of coplanar encounters. 

The cases with labels starting with ``Vel'' refer to two series of runs used to examine
the constraints due to the initial relative velocity on disk fragmentation for prograde and
retrograde encounters.  Relative encounter velocities in the range of
0.8 $km\,s^{-1}$ to 6.0 $km\,s^{-1}$ were used in the simulations to
investigate the effects of relative velocity on disk fragmentation.
The inferred disk velocities are consistent with the typical velocity
dispersion in stellar clusters \citep[a few $km\,s^{-1}$, See
e.g.,][and references therein]{cluster_bate}

\begin{table*}
\begin{minipage}{120mm}
\caption{Simulation Parameters}
\label{list}
\begin{tabular}{@{}lccccll}
\hline
Case &  Toomre $Q_{\rm min}$ & $(\theta_{1}, \phi_{1})$ (deg) & $(\theta_{2},
\phi_{2})$ (deg) & v (km/s) & ${N_{1}}$ & ${N_{2}}$ \\
\hline
 Q1 (Vel\_retro1) &   1.6   & $(180,0)$ & $(180,0)$  & 0.8 & $3\ (1)$ & $0\ (4)$  \\
 Q2  &   1.8   & $(180,0)$        & $(180,0)$       & 0.8 & $0$ &  $0\ (1)$ \\
 Q3  &   2.1   & $(180,0)$        & $(180,0)$       & 0.8 & $0$ &  $0$\\
Conf1 &  1.6 & $(180, 0)$  & $(180, 0)$ & 2.0 & $0\ (1)$ & $0\ (1)$\\
Conf2 &  1.6 & $(0, 0)$  & $(0, 0)$ & 2.0 & $3$ & $4$ \\
Conf3 (Vel\_pro2) &  1.6 & $(0, 0)$  & $(45, 0)$ & 2.0 & $1$ & $2$\\
Conf4 &  1.6 & $(180, 0)$  & $(135, 0)$ & 2.0 & $1$ & $0\ (1)$ \\
Conf5 &  1.6 & $(180, 0)$  & $(45, 0)$ & 2.0 & $2\ (1)$ & $0$ \\
Conf6 &   1.6 & $(150, 110)$ & $(150, 110)$ & 2.0 & $1\ (1)$ & $2\ (1)$ \\
Conf7 &   1.6 & $(150, 110)$ & $(150, 250)$ & 2.0 & $0$ & $4$ \\
Conf8 &   1.6 & $(150, 110)$ & $(150, 290)$ & 2.0 & $0$ & $3\ (1)$ \\
Conf9 &   1.6 & $(150, 70)$ & $(150, 70)$ & 2.0 & $2$ & $3$\\
Conf10 &   1.6 & $(150, 70)$ & $(150, 110)$ & 2.0 & $0$ & $3\ (1)$\\
Conf11 &   1.6 & $(150, 70)$ & $(150, 250)$ & 2.0 & $4$ & $3\ (1)$ \\
Conf12 &   1.6 & $(150, 70)$ & $(150, 290)$ & 2.0 & $1\ (1)$ & $2\ (1)$\\
Conf13 &   1.6 & $(30, 110)$ & $(150, 110)$ & 2.0 & $0$ & $9\ (1)$\\
Conf14 &   1.6 & $(30, 110)$ & $(150, 250)$ & 2.0 & $1\ (1)$ & $0\ (1)$ \\
Conf15 &   1.6 & $(30, 110)$ & $(150, 290)$ & 2.0 & $0$ & $5\ (1)$ \\
Conf16 &   1.6 & $(30, 70)$ & $(150, 70)$ & 2.0 & $0$ & $3\ (2)$\\
Conf17 &   1.6 & $(30, 70)$ & $(150, 110)$ & 2.0 & $0$ & $6\ (1)$ \\
Conf18 &   1.6 & $(30, 70)$ & $(150, 250)$ & 2.0 & $0$ & $0$ \\
Conf19 &   1.6 & $(30, 70)$ & $(150, 290)$ & 2.0 & $0$ & $3\ (2)$ \\
Conf20 &   1.6 & $(30, 110)$ & $(30, 110)$ & 2.0 & $3\ (1)$ & $0$ \\
Conf21 &   1.6 & $(30, 110)$ & $(30, 250)$ & 2.0 & $0$ & $0$ \\
Conf22 &   1.6 & $(30, 110)$ & $(30, 290)$ & 2.0 & $3$ & $2$ \\
Conf23 &   1.6 & $(30, 70)$ & $(30, 70)$ & 2.0  & $1\ (1)$ & $1\ (1)$\\
Conf24 &   1.6 & $(30, 70)$ & $(30, 110)$ & 2.0 & $0$ & $0$ \\
Conf25 &   1.6 & $(30, 70)$ & $(30, 250)$ & 2.0 & $0$ & $3$\\
Conf26 &   1.6 & $(30, 70)$ & $(30, 290)$ & 2.0 & $1$ & $0$\\
Conf27 &   1.6 & $(30, 110)$ & $(90, 135)$ & 2.0 & $0$ & $2$ \\
Conf28 &   1.6 & $(30, 110)$ & $(90, 225)$ & 2.0 & $0$ & $1\ (1)$\\
Conf29 &   1.6 & $(30, 110)$ & $(90, 315)$ & 2.0 & $0$ & $0$ \\
Conf30 &   1.6 & $(30, 70)$ & $(90, 45)$ & 2.0 & $0$ & $3\ (1)$ \\
Conf31 &   1.6 & $(30, 70)$ & $(90, 135)$ & 2.0 & $0$ & $5$ \\
Conf32 &   1.6 & $(30, 70)$ & $(90, 225)$ & 2.0 & $0$ & $1$ \\
Conf33 &   1.6 & $(30, 70)$ & $(90, 315)$ & 2.0 & $0$ & $3$ \\
Conf34 &   1.6 & $(90, 135)$ & $(150, 110)$ & 2.0 & $0$ & $4\ (5)$ \\
Conf35 &   1.6 & $(90, 135)$ & $(150, 250)$ & 2.0 & $5\ (1)$ & $5$ \\
Conf36 &   1.6 & $(90, 135)$ & $(150, 290)$ & 2.0 & $3\ (3)$ & $0$ \\
Conf37 &   1.6 & $(90, 45)$ & $(150, 70)$ & 2.0 & $0$ & $2\ (1)$ \\
Conf38 &   1.6 & $(90, 45)$ & $(150, 110)$ & 2.0 & $2$ & $3$ \\
Conf39 &   1.6 & $(90, 45)$ & $(150, 250)$ & 2.0 & $4\ (1)$ & $0$ \\
Conf40 &   1.6 & $(90, 45)$ & $(150, 290)$ & 2.0 & $0$ & $5$ \\
Conf41 &   1.6 & $(90, 135)$ & $(90, 135)$ & 2.0 & $0$ & $0$ \\
Conf42 &   1.6 & $(90, 135)$ & $(90, 225)$ & 2.0 & $2\ (1)$ & $0$ \\
Conf43 &   1.6 & $(90, 135)$ & $(90, 315)$ & 2.0 & $0$ & $0$ \\
Conf44 &   1.6 & $(90, 45)$ & $(90, 45)$ & 2.0 & $0$ & $0$ \\
Conf45 &   1.6 & $(90, 45)$ & $(90, 135)$ & 2.0 & $0$ & $0$ \\
Conf46 &   1.6 & $(90, 45)$ & $(90, 225)$ & 2.0 & $0$ & $0$ \\
Conf47 &   1.6 & $(90, 45)$ & $(90, 315)$ & 2.0 & $1$ & $4$ \\
Conf48 &   1.6 & $(0, 0)$ & $(90, 0)$ & 2.0 & $0$ & $3\ (1)$ \\
Vel\_pro1  &  1.6 &  $ (0,0)$     & $(45,0)$  & 0.8 & $0\ (1)$ & $0$ \\
Vel\_pro2(Conf3)  &  1.6 &  $ (0,0)$   & $(45,0)$  & 2.0 & $1$ & $2$ \\ 
Vel\_pro3  &  1.6 &  $ (0,0)$   & $(45,0)$  & 4.0 & $0$ & $0$ \\ 
Vel\_pro4  &  1.6 &  $ (0,0)$   & $(45,0)$  & 6.0 & $0$ & $0$ \\ 
Vel\_retro1(Q1)  & 1.6 &  $ (180,0)$     & $(180,0)$  & 0.8& $3\ (1)$ &$0\ (4)$ \\
Vel\_retro2  &  1.6 &  $ (180,0)$     & $(180,0)$  & 2.0 & $0\ (1)$ &$0\ (1)$ \\ 
Vel\_retro3  &  1.6 &  $ (180,0)$     & $(180,0)$  & 4.0 & $1\ (1)$ &$0$\\ 
Vel\_retro4  &  1.6 &  $ (180,0)$     & $(180,0)$  & 6.0 & $0$ & $0$ \\ 

\hline 
\end{tabular}

\medskip The pairs 
($\theta_{1}$, $\phi_{1}$) and
($\theta_{2}$, $\phi_{2}$) denote the angles between the disk angular
momentum vector and the angular momentum vector of the orbit and the
angular momentum vector and the vector direction of the initial
velocity difference in the orbital plane for disk 1 and disk 2
respectively.  $v$ denotes the relative encounter velocity.  The
number of objects formed for each disk are shown in the last two
columns.  If objects were formed within 50 AU then the numbers of
these are shown in parentheses.
\end{minipage}
\end{table*}

\section{Numerical Results of the Parameter Space Study}
\label{results}
\subsection{Varying Toomre Q parameters}
\label{toomreq}

\begin{figure*}
\includegraphics[width=\textwidth]{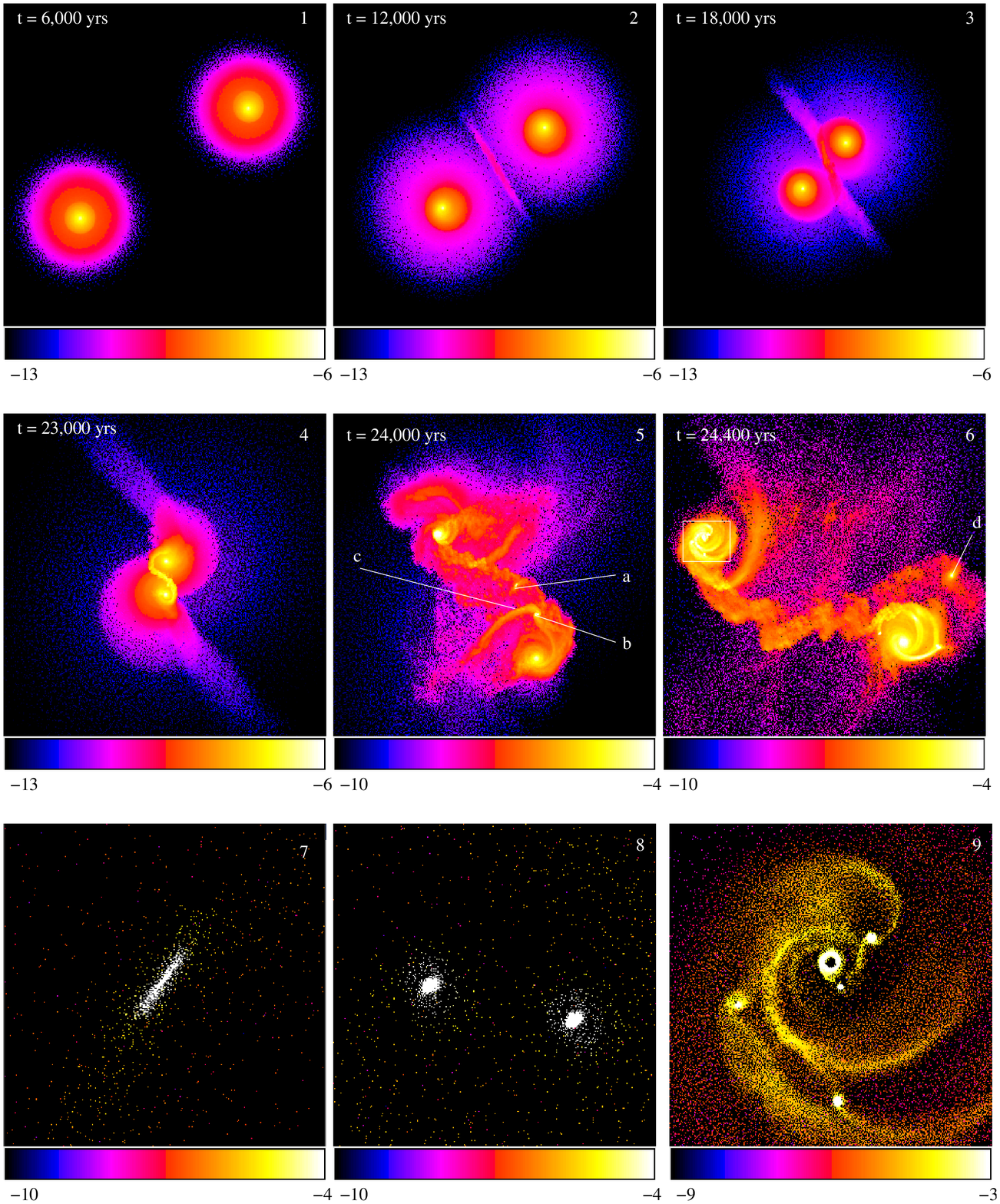}
\caption{A coplanar encounter between two retrograde
disks with Q$_{min}$ = 1.6 . Panels 1 to 6 are snapshots of the simulation, with the time
indicated at the left top of each panel.  The disk were rotating clockwise
in all of the panels. The color bar under each
panel gives the logarithm (base 10) of the density in units of $M_{\odot}/AU^{3}$.
In panels 4 and 5 only the inner dense regions were
included and a different density range was used as indicated. Panels 7 and 8 are
zoomed-in pictures of objects formed from a shock layer and due to disk
fragmentation (a and b in panel 5), respectively. Panel 9 is the
zoomed-in picture of the boxed region in panel 6, showing
fragmentation of the inner disk.}
\label{Q16_run}
\end{figure*}   

To explore the impact of the initial minimum Toomre parameter we
investigated 3 cases, all of which have $(\theta, \phi) = (180, 0)$
and relative $v = 0.8 $ km/s. These systems are only marginally
unbound. These collision parameters were chosen to ensure a strong
interaction (detailed discussion in Section~\ref{config}) so that if
no fragments form in these calculations then they are also not likely
under other configurations and encounter speeds. The two disk are
coplanar since they have the same $\theta$ and $\phi$.

\begin{figure}
\includegraphics[width = \columnwidth]{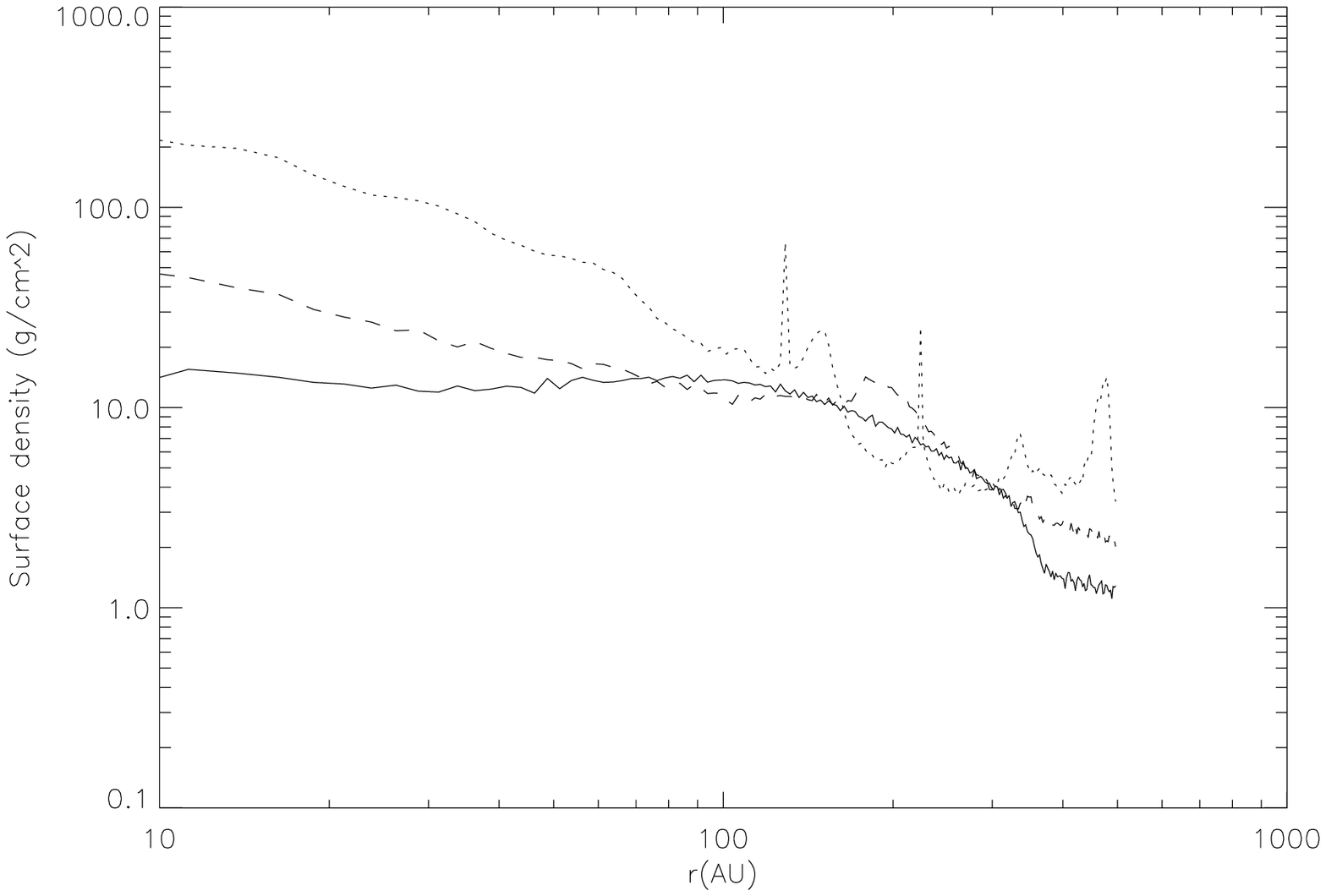}
\caption{Evolution of the surface density profile for the case in Figure~\ref{Q16_run}. \emph{Solid line}:
  Surface density profile of one of the encountering disks at 6,000
  years. \emph{Dashed line}: Surface density profile for the same disk
  at 23,000 years. \emph{Dotted line}: Surface density profile at
  24,000 years. The sharp peak at 130 AU is produced by two close objects forming
  due to disk fragmentation, the one at 225 AU originates from shock layer
  fragmentation. The moderate peak in between indicates 
  ongoing gas condensation. }
\label{sig_evolve}
\end{figure}

\begin{figure}
\includegraphics[width = \columnwidth]{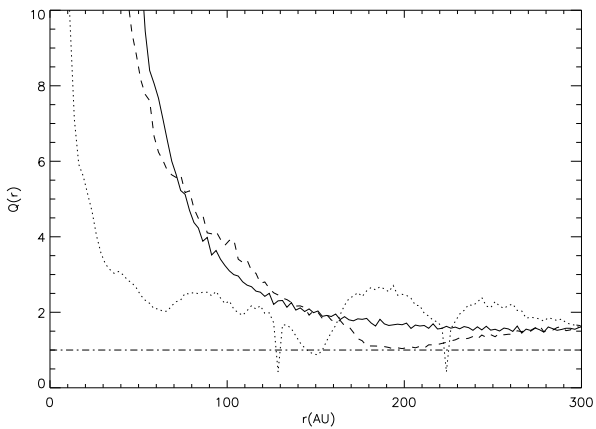}
\caption{The Toomre Q profile for the same disk as in Figure
  \ref{sig_evolve} at 6,000, 23,000 and 24,000 years (solid, dashed,
  dotted line, respectively). The dot-dashed line indicated Q = 1.0}
\label{q_evolve}
\end{figure}

Case ``Q1'' (Q$_{\rm min}$=1.6) is depicted in Figure
\ref{Q16_run}. When the physical impact began, the gas was
shock compressed and quickly formed a ``shock layer'' between the two
disks where the density was substantially enhanced.
 At 24,000 years after the encounter, the shock
fragmented to form a 7 $M_{J}$ clump at about 223 AU away from the
nearest star (label ``a'' in Panel 5, Panel 7). Also, due to its retrograde configuration, the 
angular momentum of the disk spin and the orbit are opposed and thus
the rotation speed of the gas decreases.  As a result, a large amount of gas
spirals into the inner disk (within $\sim$ 100 AU in this case), while
gas in the outer disk dissipates.  The azimuthally averaged
surface density profile of the lower-left disk in Panel 4 of
Figure~\ref{Q16_run} is shown at
different times in Figure \ref{sig_evolve} and the
corresponding evolution of azimuthally averaged Q profile is shown in Figure
\ref{q_evolve}. At t = 23,000 years, the surface density had
increased significantly, and at $r \sim 10-20\ AU$, it had almost doubled
its value from earlier times (t = 6,000 years, well before the physical impact of gas
materials). However, Q was still very high in the 
inner 100 AU because of the fixed high temperature in that region
(Figure \ref{q_evolve}).
Thus it was the presence of the shock layer at around 200 AU
that locally increased the surface density and lowered the Toomre Q
down to unity ($Q_{\rm min}$ = 1, Figure \ref{q_evolve}) allowing
a strong spiral feature to form which is gravitationally unstable.  
At t = 24,000 years, a pair of close clumps formed at r
$\sim$ 130 AU with masses of about 12 and 14 $M_{J}$ (labeled ``b'' in
Panel 5, Panel 8), and at about 150
AU another gas blob started condensing (a moderate extremum near the
130 AU clump in the plot of the Q profiles, labeled ``c'' in Panel 5), 
which eventually became a
bound object (labeled ``d'' in Panel 6). Gas inflow increased after formation of these
objects. At t = 24,400 years,
fragmentation within 20 AU was finally triggered in one of the
disks and 4 brown-dwarf mass clumps were formed (boxed region in Panel 6,
Panel 9).  However, these objects were not included in our 
statistics (Section \ref{objects}) because the
density within 20 AU at this time step exceeded $10^{-12} g \ cm^{-3}$ and became
optically thick (cf. Section \ref{cooling}), so that an isothermal cooling
assumption was no longer appropriate.  The simulation was stopped at t =
24,400 years because the time step reduced dramatically after the
first condensation formed.  Analysis based on the energy and momentum
at the end of simulation indicates that amongst the 4 objects formed
outside  $100 \ AU$, one is unbound and likely to leave the system, one
has a large semi-major axis (a = 879.3 AU) and the other two have
orbits with semi-major axes less than 100 AU.                

The encounter ``Q2'' has same parameters as ``Q1'' except the
initial $Q_{\rm min} = 1.8$. The disk underwent similar physical
processes such as formation of a layer of shock-compressed gas, 
inflow of gas and disk truncation after the
encounter. Nevertheless, the disks were more stable to the perturbation
and only one of the disks developed an m = 2 mode gravitational instability
at a relatively late time step (t = 25,300 years), which then
fragmented and formed only one bound object at radius of 30
AU. The other disk 
was truncated but remained stable. The gas in the shock layer
did not condense to form an object.  Instead, it
dispersed after the disks moved apart.  The lowest Q value
at t = 24600 years (the time step just before fragmentation) for the
unstable disk was 1.1 and for the stable disk was 1.7, close to the original minimum.

The last case in this series ``Q3'' used initial
disks with $Q_{\rm min}$ = 2.1. The encounter disrupted
both disks without forming any objects.  The two stars
happened to capture each other and formed a close binary system with
a separation of about 20 AU.  Most of gas was dispersed, with the
bound gas at the end of simulation forming a circumbinary disk with
a radius of 250 AU at 31000 years. The
circumbinary disk had a much lower surface density compared to the initial disks. 
Assuming the gas was still in
Keplerian rotation, the minimum Toomre Q was very high ($Q_{\rm min} =
6.2$). The disk was thus very stable against spontaneous
fragmentation and no clumps were expected to form in future evolution.    
  
The cases described here showed that instability and
fragmentation can be triggered in an initially stable disk by pairwise
encounters. However, the initial minimum Toomre parameter can not be
higher than 2.1 for instabilities to occur.  From the evolution of the
Q profile in Figure \ref{q_evolve} it can be seen that lowering Q down to
unity is crucial for fragmentation.  Two competing processes,
gas inflow and dispersion, proceed at the same time in an
encountering disk.  In a high Q disk the gas is hotter and
an encounter is more likely to accelerate the gas dispersion rather
than triggering gas inflow as in the relatively low Q (but still
initially stable) cases.  Indications are that early stage proto-stellar disks are
usually large, heavy and relatively cold, with temperatures not far
from those of the molecular cloud core ($\sim 10 K$).  So gas inflow can be the
key mechanism in encounter-induced disk fragmentation. 

Note that the local enhancement of density in the shocked gas can
lower the Q parameter significantly and induce disk instability before the
inflowing gas reduces Q in the inner, hotter region, 
making the fragments form relatively far from the stars. This is
especially clear for coplanar encounters discussed in 
this section.  In more likely, non-coplanar cases (cf. Section
\ref{config}) the shocked gas usually does not form a well-defined
layer between the disks, but it remains an important factor driving disk
instability.  In the case  ``Q1'' the shock
layer itself also fragmented directly.  The object formed this way was
not associated with spiral structure in the unstable disk (panel 5
in Figure \ref{Q16_run}).  Fragmentation of shock layers rarely
occurs and was restricted to coplanar, low-velocity, low-Q cases. With
higher disk temperature (as in ``Q2'') or higher velocity the
shock layer usually dissipates when the two stars pass 
periastron without forming objects.          

\subsection{Varying the Encounter Configurations}
\label{config}
The parameter space of encounter configurations was explored with 48
encounters covering the $(\theta, \phi)$ parameter space. The effect
of the relation between disks' spin and orbital angular momentum in z
direction can be see by varying the disks' $\theta$
parameters.  This variation qualitatively changes the interactions,
including the mechanisms of fragmentation.

There is a large stochastic element to the evolution of the disks and
the formation of clumps.  In this sense, the precise number of objects
formed in a single simulation should not be examined too closely.
For example, a small perturbation in the initial particle placements
that is within the noise in the initial surface density profile is
sufficient to change the outcome in terms of the precise number and
placements of clumps.  For example, several cases are symmetric from
the point of view of the two disks but have non-symmetric outcomes.
Symmetry in the configurations is revealed by the mappings $x,y,z
\rightarrow -x,-y,z$ which implies $\theta_1=\theta_2$ and $
\phi_1=\phi_2$ (such as for the coplanar cases and ``Conf6'', for
example) and $x,y,z \rightarrow -x,-y,z$ which gives
$\theta_1=\theta_2$ and $ \phi_1=\phi_2+180$ (e.g. ``Conf11'').  These
initial states were created using absolute offsets ($x \rightarrow x+\Delta x$)
of two identical disks that have small non-axisymmetric perturbations
due to the glass initial particle state.  Thus the small perturbations in
the particle configurations are not symmetric in these cases.  These
difference magnify during the course of the encounter and ultimately
result in different numbers of objects in each disk as can be seen in
the object numbers for each disk in symmetric cases in
table~\ref{list}.  For example disk 1 in ``Conf6'' made 2 objects
while Disk 2 made 3.  The variations are consistent with a
Poisson-type random process.   In most of our analyses we look at outcomes
averaged over several encounters.

\subsubsection{Formation of Tidal Structures in Prograde Disks and
  Fragmentation} 

\begin{figure*}
\includegraphics[width=\textwidth]{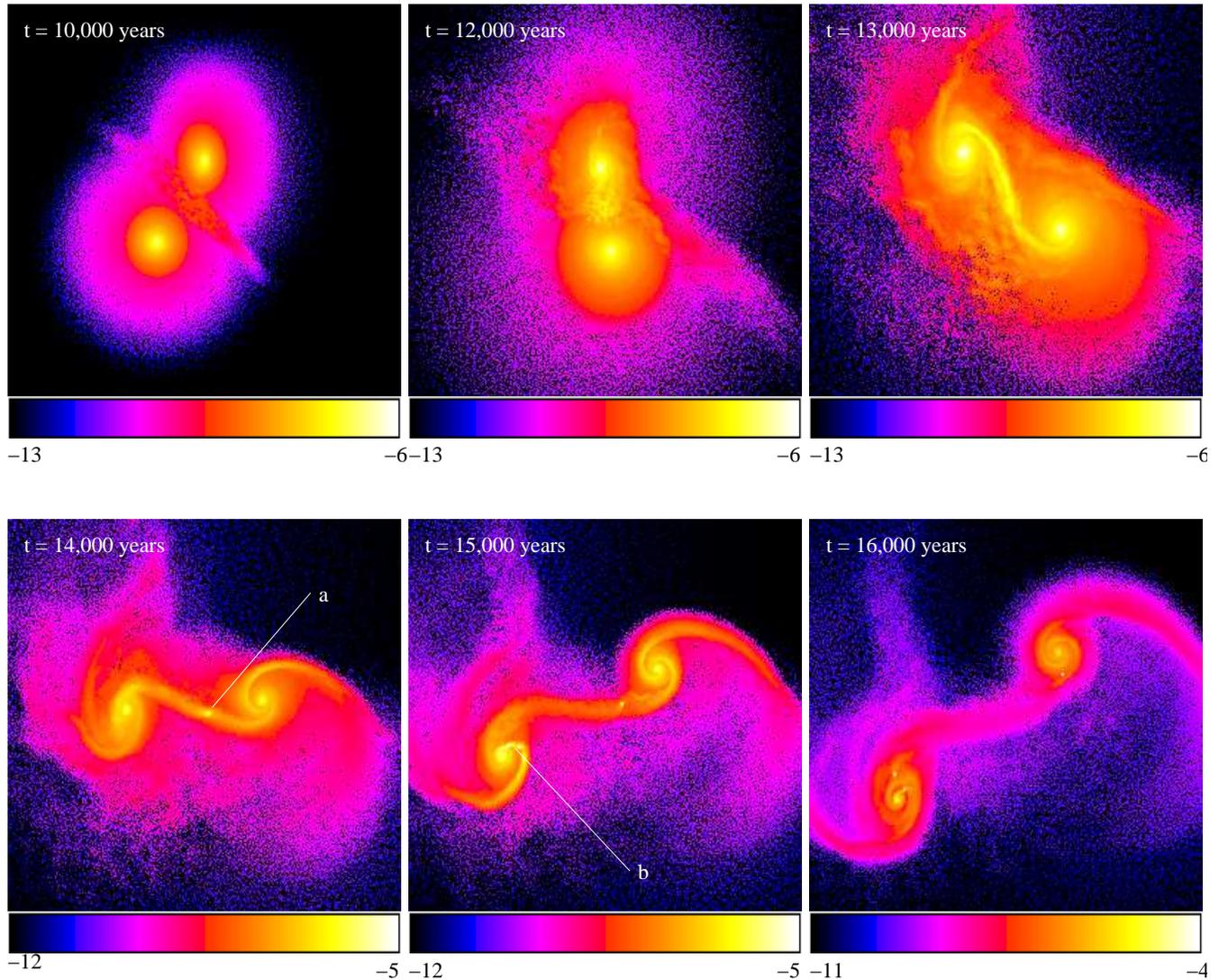}
\caption{Snapshots of the prograde-prograde disk encounter ``Conf3''. Time
  is indicated on the left top of each panel. The disks are counter-clockwise
  rotating in this view. The color bar under each
  panel gives the logarithm (based 10) of the density in units of
  $M_{\odot}/AU^{3}$.  Different density ranges are used to clearly
  show the clumps.  }
\label{pp45}
\end{figure*}

Encounters between retrograde disks tended to reduce the disk spins and
cause gas inflows, increasing the surface density of the inner
disks and reducing the Toomre Q parameter (Figure
\ref{Q16_run}).  Prograde encounters are more complex as seen in
Figure \ref{pp45} showing snapshots of encounter ``Conf3''.  The two disks have $(\theta,
\phi) = (0, 0)$ and $(45,0)$, respectively.  
The encounter had a relative speed 2.0 km/s and the initial minimum Q of the
disks was 1.6.  Gas between the two disks was shock-compressed
(Panel 2 in Figure \ref{pp45}), though it did not form a layer or
fragment.  In prograde cases spin-orbit
resonances can occur.  Gas in the
resonance region is continuously perturbed promoting angular
momentum transport that simultaneously enhances the density of the inner disk
while the outward gas flows collect a substantial portion of the disk mass into
filamentary tidal structures (Figure \ref{pp45}, Panel
4), analogous to the tidal tail structures generated during the interaction between two
prograde disk galaxies.  In the case shown, at 14,000 years gas 
began condensing from a tidal structure of one disk (on the right in
Panel 4) at about 320 AU from the star (labeled ``a'' in
Panel 4), which later formed a bound object of mass $\sim 20 \
M_{J}$. A tidal structure was 
also present in the other disk, where there was a 45 degree angle
between the spin and orbital angular momentum vectors.
At 15,000 years, it condensed into two objects with masses of 9
$M_{J}$ and 70 $M_{J}$ at 60 and 110 AU from the star, respectively
(labeled ``b'' in Panel 5 Figure \ref{pp45}). The simulation was
terminated at 16,000 years.  Analysis based on the energy and
momentum of the objects found that all the 3 objects remained bound to the
system but their orbits were large: 532 AU for the one that condensed
first, and 90 and 307 AU for the two condensed in the later time
step. 

Fragmentation of tidal structures was seen in several prograde
disk encounters, especially when the disk was closely aligned with the orbital
plane.  This result is consistent with \citet{lin}, in which a
disk-disk encounter triggered formation of a large tidal tail and an
object condensed from it.  Objects formed this way tend to have large
orbital radii.  For example, in the case``Conf2'', tidal structures from both disks 
condensed and formed 7 objects in total by the end of simulation, all
of which were unbound to both stars.  

\subsubsection{Prograde vs. Retrograde Encounters}

Tidal structures due to spin-orbit resonances are not as efficient at
producing objects as the inner disk fragmentation seen in retrograde
disks.  On the other hand, \cite{watkins} has argued 
that star formation triggered by collisions of molecular
clouds tends to result in prograde-prograde coplanar encounters so
they may be more common. 

\begin{figure}
\includegraphics[width=\columnwidth]{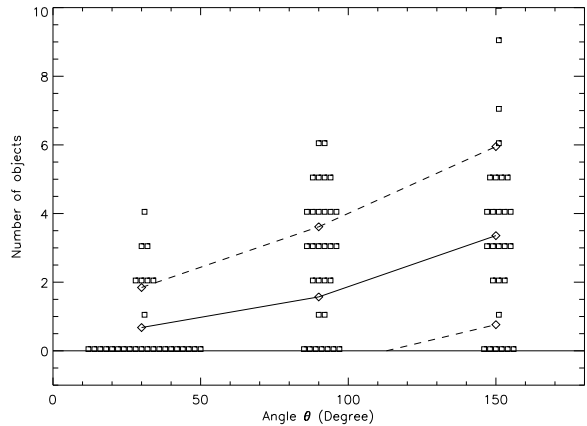}
\caption{The number of objects formed in a disk per encounter as a
  function of $\theta$.   Each box corresponds to a single simulated disk that
  produced the number of objects indicated on the y-axis.  The boxes
  have been spaced apart in the x-direction for clarity.  The average
  and variance are indicated with solid and dashed lines respectively.}
\label{thetatrend}
\end{figure}

In Figure \ref{thetatrend} we plot the
number of object formed in a disk per encounter as a function of $\theta$,
the angle between the individual disk spin vector and the positive z
direction. We chose the homogeneous sample from ``Conf6'' to ``Conf48'' in
Table \ref{list}, excluding the unlikely coplanar
encounters, $\phi=0$ encounters and encounters with disks residing in
the orbit plane.   
The cases chosen all have same initial Toomre
parameter and relative encounter speed.  

The average number of bound and unbound objects formed within that disk
increases with increasing $\theta$ for that disk.  This is the case regardless of the
configuration of the other disk in the interaction.  Gas inflow 
efficiently adds mass to inner disks which fragment whereas in prograde disks, 
the mass transfer usually send substantial amounts of gas outward
rather than inward.  The condensation of tidal structures did not occur
frequently, because in most of the cases they usually dissipated before
anything could form.         

\subsubsection{Plane-Parallel Encounters }

No clear correlation was found between the productivity of a disk
and the disk's own $\phi$ parameter.  However, no
fragmentation would happen if the two disk planes are parallel
(or nearly parallel) initially but not coplanar. 60\% of the encounters in the
``Conf'' series which do not produce any clumps fall into this
category. The disks were simply truncated in these
cases. The final size of the disk is about 300 to 500 AU depending on
the initial configuration. 

\subsection{Varying Encounter Velocities}

We investigated the effect of varying the relative velocity with two series
of simulations, one containing non-coplanar encounters with both disks
prograde, and one with coplanar encounters with two
retrograde disks.  In each series, the relative velocity of the
encounters is repeated along with the number of object formed in Table~\ref{velocity}. 

\begin{table}
\caption{Number of objects formed in the simulations with different
  encountering velocities}
\label{velocity}
\begin{tabular}{@{}cccc}
\hline
Case & Velocity & Number & Fragmentation  \\
     & (km/s)  & of Clumps & Mechanism \\
\hline
Vel\_pro1 & 0.8 & 1 & Circumbinary disk \\
Vel\_pro2 & 2.0 & 3 & Tidal structure and disk \\
Vel\_pro3 & 4.0 & 0 &  \\
Vel\_pro4 & 6.0 & 0 &  \\
Vel\_retro1 & 0.8 & 8 & Shock layer and disk  \\
Vel\_retro2 & 2.0 & 2 &  Disk only \\ 
Vel\_retro3 & 4.0 & 3 &  Disk only \\ 
Vel\_retro4 & 6.0 & 1 &  \\ 

\hline
\end{tabular}
\end{table} 

With the exception of a special single case (``Vel\_pro1'', see
below), the number of clumps formed in 
an encounter generally decreases with increasing velocity.  No
clumps formed in prograde disks when the relative
encounter velocity was above 4.0 km/s.  This result can be
qualitatively understood by 
comparing the interaction timescale with the dynamical timescale of the
disks.  Though gravitational interactions occur regardless
whether or not there is a physical impact, these simulations indicate that
significant changes in the density profile occur only when the
disks are close enough to physically impact.  The interaction
timescale is approximately, 
\begin{eqnarray}
t_{\rm int} = \frac{d_{\rm eff}}{v}
\label{tenc}
\end{eqnarray} 
where $d_{\rm eff}$ is the twice the disk diameter (4000 AU in this
case), which gives the rough separation for physical impact. $v$ is
the relative velocity of the encounter. The dynamical timescale within
each disk is a function of radius $r$, given by,
\begin{eqnarray}
t_{\rm dyn}(r) = \frac{2\pi r^{3/2}}{\sqrt{GM_{\ast}}}
\end{eqnarray}   
where $G$ is the gravitational constant and $M_{\ast}$ is the mass of the
star.   The encounter timescales $t_{\rm int}$ were $2.4
\times 10^{4}$, $9.5 \times 10^{3}$, $4.7 \times 10^{3}$ and $3.1
\times 10^{3}$ years for relative speeds 0.8, 2.0, 4.0 and 6.0 km/s,
respectively.  The dynamical timescale $t_{\rm dyn}$ was about $1.1
\times 10^{4}$ years at 400 AU and $3.9 \times 10^{3}$ year at 200 AU.
The implication is that there was insufficient time during fast encounters,
where $t_{\rm int} < t_{\rm dyn}$ to apply the torque that transfers
angular momentum to change the disk profiles and causes fragmentation.
This is more restrictive for prograde disks where tidal structures form
further out at distances similar to 400 AU.

\subsection{Captures and the Formation of Binary Stellar Systems}

In the case ``Vel\_pro1'', star capture occurred and the two 
encountering stars formed a binary system. Both
disks were prograde and large tidal tail structures
formed when the disks passed periastron.  The tidal structure
dissipated over time and carried away the excess angular momentum from
the system, enhancing the 
formation of a close binary.  The separation of the two components was
about 90 AU, significantly smaller than the periastron of the
trajectory (about 300 AU taking into account the effect of
gravitational focusing). The circumbinary disk was about 500 AU in
radius.  At a later time it also fragmented to form a 20 Jupiter
mass object.
 
Another case where stellar capture occurred is the case ``Q3'' (as described in Section
\ref{toomreq}).  Since the starting disk Q
parameter is 2.1, the circumbinary disk became hotter and less massive
compared to the standard cases.  No fragments formed at the end of the
simulation.  The initial disks are retrograde and the encounter did not induce tidal
structures. 
 
The two cases above are the only ones where stellar capture occurred. 
Note that both of them have rather
conservative choices of encounter velocities (0.8 km/s), with which the system are
marginally bound at the initial state. 

\section{Properties of the Objects Formed} 
\label{objects}

Most of the encounters produced clumps with the exception of those
with high relative velocities or very stable initial disks
(e.g. $Q_{\rm min}$ $>$ 2.1).  As described in the last section, bound
objects can form from fragmentation of shock layers, instabilities
within a disk and in tidal structures.  Since the timescale of disk
encounters (about 10,000 years) is short and encounters may happen
when there is still a lot of material obscuring the stars, encounters in
progress are unlikely to be observed whereas the sub-stellar objects
formed may persist indefinitely.  Though these objects may evolve
substantially from their time of formation, particularly in terms of
luminosity, simulated clump properties such as mass, angular momentum
and orbital kinematics can be directly compared with properties of
observed objects.  In the initial state, the disks were well resolved
outside 50 AU.  However, as the simulations progressed, additional
matter accumulated in the inner disks improving the resolution
elements per scale height.  On the other hand, objects with 50 AU are
likely to experience further evolution such as accretion and migration
into the star.  In addition, work by \cite{rafikov} has suggested that
disks this close to a star are optically thick and unlikely to cool
fast enough to fragment.  For these reasons, we have excluded objects
within 50 AU from the sample discussed below.  However, the objects
formed within 50 AU are not statistically different and properties
such as the relative mass distribution would not be significantly
altered if these objects were included.  A total of 191 objects formed
in our simulations (excluding the high Q cases) with 144 outside 50 AU.

\subsection{Mass Distribution} 

\begin{figure}
\includegraphics[width=\columnwidth]{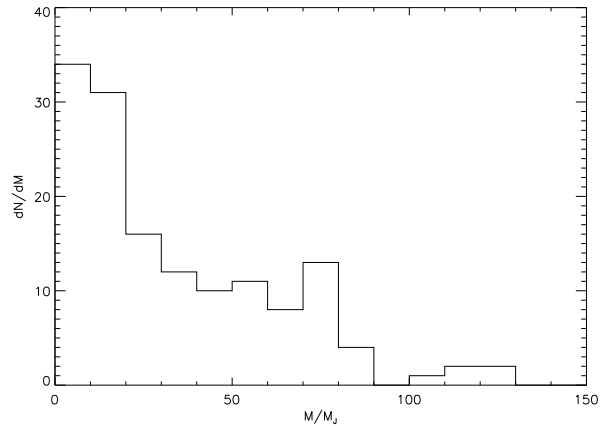}
\caption{The mass distribution of 146 objects formed outside 50
  AU in the 48 ``Conf'' simulations. The bin size is 5 $M_{\rm J}$.
  }
\label{mass}
\end{figure}

The object masses range from 0.9 Jupiter masses ($M_{\rm J}$) to
127 $M_{\rm J}$, extending into the stellar regime. The mass distribution is
plotted in Figure \ref{mass}.  It was found that the number of
objects generally decreases with increasing mass, consistent with the
observed sub-stellar initial mass function (IMF).  The simulated
population does not include objects much below 10 $M_{\rm J}$). 
In this regime there are 1000 or fewer particles per object.   The numerical
gravitational softening (0.2 AU) could also inhibit the formation of
smaller objects.  Thus these simulations are unable to probe whether
there is a physical mass cut-off in this range.

\begin{figure}
\includegraphics[width = \columnwidth]{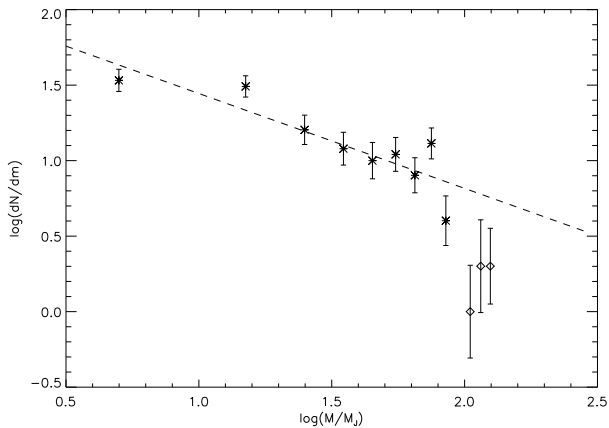}
\caption{The logarithmic mass distribution of the objects. The error
  bars were calculated assuming Poisson errors $\protect\propto
  \sqrt{N_{\rm bin}}$. {\it Stars}: sub-stellar objects ($M \protect\lid 80
  M_{J}$); {\it Diamonds}: low mass stars ($M > 80 M_{J}$); {\it Dashed
   line}: linear fit of the logarithmic distribution of sub-stellar
  objects $dN/dM \protect\propto M^{-\protect\alpha}$,
  with \protect$\alpha = 0.6$}
\label{logmass}
\end{figure}

Observationally, the sub-stellar IMF in open
clusters can be well described by a power-law $dN/dM \propto
M^{-\alpha}$.  For young open clusters such as Pleiades,
$\lambda$ Orionis and $\sigma$ Orionis, the exponent $\alpha$ is close
to 0.6 at low masses with an uncertainty of order 0.1
\citep{moraux,barrado,caballero}. An universal IMF was proposed by
\citep{kroupa} in which $\alpha = 0.3 \pm 0.7$ in the range
$0.01M_{\sun} < M_{\ast} < 0.08 M_{\sun}$.  In this work, we fit the
mass distribution of the sub-stellar region ($M < 0.08 M_{\sun}$) and
the best fit gives $\alpha = 0.6$ with $\chi^{2}$ uncertainty 0.15
(Figure \ref{logmass}), consistent with the observed values.  Above
the stellar boundary the number of objects declines more
rapidly with mass and a break is apparent near 0.08 $M_{\sun}$ (or about 0.1 of
the central mass) which is also consistent with the trend reported by
\citep{kroupa}.  There are only 9 objects in total that are
beyond 0.08 $M_{\sun}$ and thus the statistics are poor.  More
simulations, including those with varying central star masses, are required to
explore the resulting mass function in the low mass, stellar regime.

\subsection{Shapes and Angular Momenta} 

\begin{figure}
\includegraphics[width=\columnwidth]{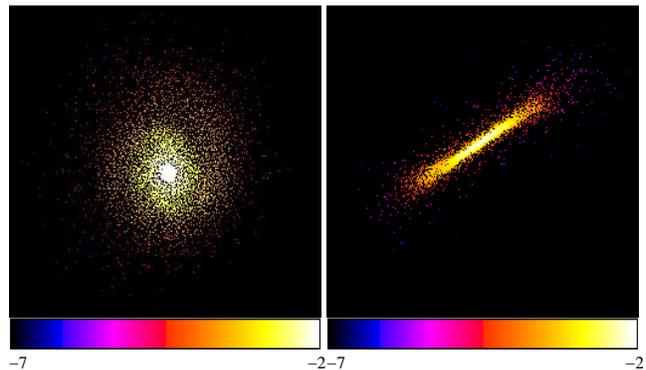}
\caption{A resultant clump with typical disky shape.  Both panels are
  14 AU on a side.  The clump has disk radius about 4
  AU. {\it Left}: The object projected in the x-y plane;
  {\it Right}: The object projected in x-z plane. The gray scale
  indicates the density in logarithmic space in code units $
  M_{\sun}/AU^{-3}$.}
\label{clump}
\end{figure}

Most of resultant clumps in our simulation are not spherical but have
highly flattened, disky shapes (height to radius ratio $\sim$ 0.1), as
shown in Figure~\ref{clump}.  The radii of the 
disks range from 0.1 AU to 10.0 AU.  The lower bound is
likely to be associated with the resolution limit (0.2
AU). The disky shape is maintained by the high spin angular
momentum of each clump.  Depending on the mass and size, the total
angular momenta of the clumps range from $10^{47}$ to $10^{51}$ g
cm$^{2}$/s.  Assuming the total angular momentum is conserved during the
evolution and adopting a typical radius of a brown dwarf to be 
$7.0 \times 10^{9}$ cm \citep{burrows}, the rotation speed at the edge
of the brown dwarf would be 10 times the break up speed.  Thus the
proto-brown dwarfs disks must shed rotation in order to
collapse down to a proto-BD.  This result is consistent with recent
mid-IR, sub-millimeter, and millimeter observations of the spectral energy
distributions (SEDs) of young brown dwarfs, which indicate that most of
them are surrounded by circumstellar disks \citep[e.g.,][]{muench}.  A
T Tauri accretion phase has also been indicated in some young BD disks
by the detection of broad, asymmetrical $H\alpha$ lines
\citep{jay}.  The IR excess in SEDs was found to decrease as the age of
BDs increases, which suggests that the BD disks evolve from flared
to flat geometry similar to proto-stellar disks \citep{mohanty}. The
clumps in our simulations are flat partly due to the assumed
isothermal EOS.   As explored in section~\ref{cooling}, these dense
condensations are probably optically thick and would thus be
hotter and more puffed up in nature.
                  
The simulations were terminated a few thousand years after first object
formed due to computer time limitations. 
In some objects, like the one shown in Figure \ref{clump}, by
the time the simulation was halted, the gas with lower
specific angular momentum had started to condense into the center and
a density gradient had also been established.  Because the gravitational
softening (0.2 AU) is comparable to the size of most clumps,
our ability to resolve the structure and follow the evolution of the
proto-BD disks is limited.  However, it can be expected in later
evolution a central proto BD will form and the surrounding disk will
accrete onto it on a viscous timescale.    

In the majority of cases where clumps formed from fragmentation of disks,
the spin directions of the proto-BD disks are similar to their
parent proto-stellar disks.  In the case where the shock layer fragments,
however, the resultant proto-BD disk may spin in a direction different
from both disks.  For the proto-BD disks that are still orbiting the star, further
interaction with their parent disk will influence the final spin.

\subsection{Binary Brown Dwarfs and Multiple Systems} 
\label{binary}

Typically where fragmentation occurred more than one object was
 produced.  About 15 percent of the
objects are in BD-BD binary systems.  This fraction is consistent with
 the observed binary fraction (around $10\pm 10$
\% at the range 0.01 to 0.1 $M_{\sun}$) \citep[cf. Figure
3][]{hubber}.  In most cases, the binaries formed from secondary
fragmentation of very large disky clumps.  Further numerical studies
with higher resolution are necessary to investigate this question in detail
because, although the Jeans mass was well resolved before
fragmentation and most of the clumps are resolved by at least 1000
particles, the Jeans masses of the clump themselves decrease so quickly with
increasing density that they are only marginally resolved in our
simulations.  Without resolving the Jeans mass, it has been suggested
 that fragmentation could be artificially enhanced \citep{bate_res}.  However, given
the very high spin rate of the clumps, secondary fragmentation is likely to occur,
according to \citet{matsumoto}.  

Most of the BD binary systems were still bound to the central stars when the
simulations were terminated.  For example, in case ``Conf5'', the binary
system orbits the star at $\sim$ 140 AU, with the two components (mass
19 $M_{\rm J}$ and 13 $M_{\rm J}$) orbiting each other in a period of
around 30 years and with separation about 3 AU.  Each component has
a highly flattened shape. In future evolution, it is expected that some
of these binaries will merge to the star, leave the system, or be ejected
in close interactions with other objects.  
The surviving systems are likely to remain hierarchical star-BD
multiples.  Thus the collision mechanism may explain the origin of observed star-BD
multiple systems such as GL 569, where a recently confirmed BD triple
(GL 569B) orbits the primary, an M2.5V star (GL 569 A) \citep{gorlova,
simon}.  The system is about 100-125 Myrs old so no BD disks were
observed. However, in younger BD multiples accretion signals have been
detected \citep{kraus}, indicating accretion disks surrounding 
components of the multiple.  

\subsection{Orbits and Further Evolution of the Resultant Clumps} 

Objects in our simulations typically formed at least 50 AU away from
the closest star.  Approximately 60 percent of the objects formed were
not bound to the system.  The percentage is higher for coplanar
encounters, in which the shock layer fragmentation can take place
relatively far out in the gravitational potential of the stars.  The
unbound objects may leave the whole system to become free floating BDs
(or BD binaries) or low mass stars.  For the objects that are bound to
the star, the orbital properties vary from case to case.  About 5
percent of the orbits have very large semi-major axes ($>$ 500 AU) but
are still bound to the system. Since the outer proto-stellar disk was
dispersing in most encounters, sub-stellar objects in large orbits
were unlikely to accrete additional gas to become low mass
stars. Hence these objects may remain as wide-separation BD
companions, which have been detected in observations \citep{gizis}.
Where more than one object orbits the star on nearby orbits dynamical
interactions can take place resulting in the preferential ejection of
lower mass objects \citep{reipurth}.  The 11 objects that were ejected
to significant distances during the simulations were lower in mass.
However, the majority of the objects were unbound and would ultimately become
free-floating.

For the clumps that stay in the dense region of the proto-stellar disks
(within 100 AU), the future growth and evolution depend on the rate
that the gas accretes onto the objects.  Under the assumption of efficient
gas accretion, as in the competitive accretion scenario
\citep{reipurth, cluster_bate}, these objects may eventually
become low-mass stars over timescales much longer than those simulated here.
 
\section{Cooling Approximations}
\label{cooling}
\subsection{Optically Thick Approximation}
The simulated disks produced dense regions as they evolved and reached
the point where they were no longer uniformly optically thin.  The disk material
itself tended to remain optically thin ($\tau < 1$) and was
thus capable of efficient cooling but the fragments within the disk
reached high densities where
the radiative cooling times became very long.  Thus on the timescale of
these simulations, the interiors of the fragments were effectively
adiabatic.  

In earlier work on proto-planetary disk fragmentation \citep{mayer},
the approximation of converting the entire 
simulation to adiabatic gas was employed to test the behaviour in this
regime.  This corresponds to the extreme where cooling is completely
absent.  A less extreme test with strong suppressed cooling is to
smoothly interpolate from efficient black-body cooling when the gas is
optically thin to no cooling in the optically thick regime.  
Optically thin black-body
radiative losses are given by, 
\begin{eqnarray}
L = 4 \kappa \sigma T^4 M
\label{optthinemit}
\end{eqnarray}
where $L$ is the luminosity, $\kappa$ is the opacity, $T$ is the
temperature and $M$ is the total mass.  The opacity was taken to be
1 cm$^2$ g$^{-1}$ at $T=60 K$ with a power-law variation with
temperature as $T^{1.3}$.  This is consistent with the dust opacity
values calculated by d'Alessio et al., as tabulated in \citet{mejia}, over
the range of 10-300 K.  

We then took a late stage of a representative run ``Conf30'' at 15000 years, after
fragmentation had occurred, and measured the optical depth to infinity
in 12 directions (corresponding the faces of a dodecahedron) for each
particle.  This allowed us to estimate the averaged effective optical
depth, $\tau_{\rm eff} = -{\rm log}_e \int_{\Omega} e^{-\tau(\Omega)}
d\Omega / 4 \pi$.  We found that the effective optical depth correlates
fairly well with gas density in the regime, $\rho = 10^{-16}$ to
$10^{-12}$ g cm$^{-3}$, where the material transitions from optically
thin to ($\tau_{\rm eff} \sim 0.1$) to optically thick ($\tau_{\rm eff}
\sim 10$) as $\tau_{\rm eff} = 10^{8.1+ 0.56 \rho}$.  The {\it rms}
deviation of the fit is $0.16$ dec in $\tau$.  Assuming isotropic
emission, we can then estimate the fraction of the emitted light
(given by equation~\ref{optthinemit}) that escapes the disk entirely.
We then assume that the rest of the emission is absorbed on-the-spot,
an approximation commonly used for optically thick lines in models of
emission nebulae, so that the net radiative loss, $L_p$, from one
particle is $L_p = 4 \kappa \sigma T_p^4 m_p e^{-\tau_{\rm
    eff}(\rho_p)}$, where $m_p$, $T_p$ and $\rho_p$ are the particle
mass, temperature and density, respectively.  In the limit of
infinitesimal particle masses, this expression is proportional to the
contribution to the surface flux from a mass of gas, $dm =
\rho\,dl\,dA = d\tau/\kappa\,dA$.  In the case of an infinite
plane-parallel slab, when an analytic expression for $\tau_{\rm eff}$
is used and the flux contribution is integrated from the surface to
high optical depth the total surface flux asymptotically approaches
the value of $\sigma T_p^4$.

The assumption for the main set of simulations was that the disk
temperatures were set by heating from the central star offset by
radiative losses so that the net temperatures were a function
of the distance to the nearest star.  To include the effect
of heating by the star, the governing equation of the
particle energy per unit mass, $E_p$, is,
\begin{eqnarray}
\frac{d\,E_p}{dt} = 4 \kappa \sigma e^{-\tau_{\rm eff}(\rho_p)}
(T(r_{\rm star})^4 - T_p^4) - \frac{P}{\rho} \nabla.{\bf v}|_p + \Lambda_{{\rm shock},p},
\end{eqnarray}
where $T(r_{\rm star})$ is the temperature determined from an
equilibrium between stellar heating and radiative losses at the radius
$r_{\rm star}$ from the nearest star, $-\frac{P}{\rho} \nabla.{\bf v}$
is the compressive heating term and $\Lambda_{\rm shock}$ is the shock
heating term.  Thus the model now includes compressive heating and
shock heating.  Outside the intense collision environment, where these
process are unimportant in relation to stellar heating, the disk
temperatures relax to the temperature as a function of radius assumed
in the preceding sections.  Overall, this similar to the approximation
employed by \cite{stamatellos}, where local cooling was based on the
amount of radiation escaping from regions with similar physical
conditions in spherical collapse models.  In the approximation used
here the radiative loss rates were calibrated from similar disk cases.

Our approximation retains thermal energy
indefinitely in the denser gas and is thus conservative with regard to
ensuring that the fragmentation is not a result of favourable cooling
assumptions.  Detailed radiative transfer calculations should give
results between this approximation and the isothermal EOS.  It has
been argued by \citet{rafikov} that the optically thick parts of
proto-stellar disks are likely to be convective, in which case our
neglect of radiation as a heat transport mechanism may not have a
large impact on the disk structure.  However, the cooling of the
optically thin surface layers are approximately correct in this model.
Full radiative transfer has been implemented within the Gasoline code
(Rogers, Shen \& Wadsley, {\it in preparation}) and will be applied to proto-stellar disk collisions as
part of future work.

\subsection{The Optically Thick Run} 

\begin{figure}
\includegraphics[width=\columnwidth]{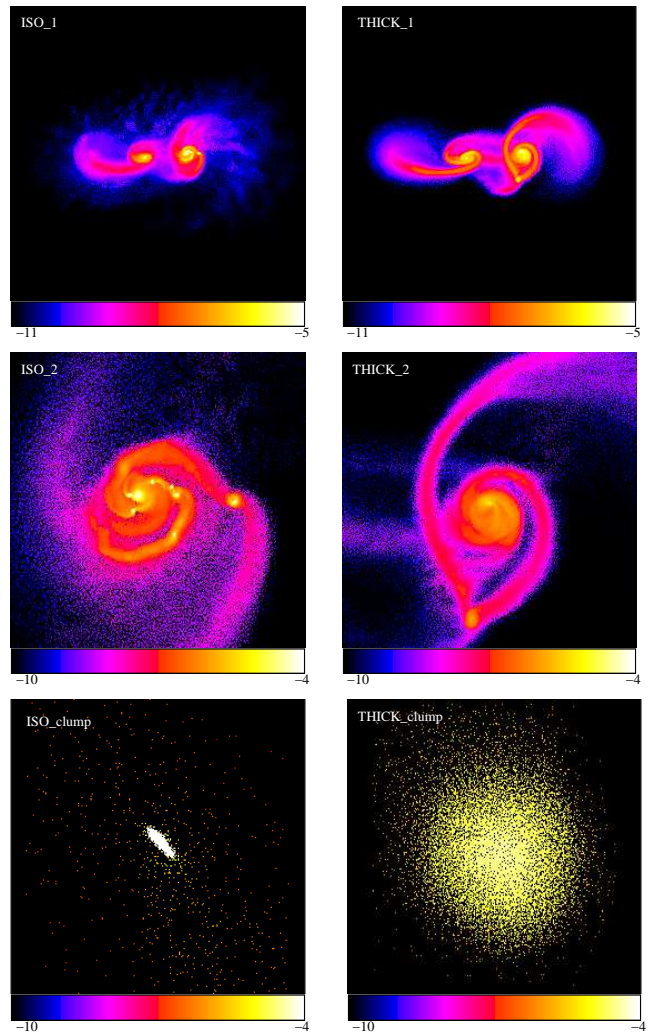}
\caption{A Q = 1.6 disk encounter with isothermal EOS vs. our optical
thick approximation. {\it Left}: The run with isothermal EOS (top and middle
panels) and the typical shape of a clump (bottom panel); {\it
  Right}:  The simulation that used optical thick approximation. 
The grey scale under each panel gives the density
ranges in units of $M_{\sun}/AU^{3}$.}
\label{optical_thick}
\end{figure}

We re-ran the simulation we used to calibrate the effective optical
depth-density relationship using the new cooling approximation. Figure
\ref{optical_thick} compares the disks at 14,000 years in the locally
isothermal run (Panels that are labeled with ``ISO\_1'' and ``ISO\_2'')
with the ones in the new run (Panels labeled with ``THICK\_1'' and
``THICK\_2''). Using the isothermal equation of state, one 
of the disks fragmented and formed ten bound objects, including
two binary systems.  However, when using the optical 
thick approximation, the cooling is suppressed and the disk does not
fragment immediately. Instead, it is 
the tidal structure induced by the gravitational resonance that condensed
and formed a clump at about 410 AU from the star. At later time steps, the
gas in the tidal tail dissipated and the objects orbit the star at about
350 AU. The clump had a mass of 0.075 $M_{\sun}$, close to the hydrogen
burning limit.

Compared to the objects that formed in the
corresponding isothermal run, the clump had a more spherical
shape.  Nevertheless, the angular momentum of the object was still very
high, about $7.5 \times 10^{50}$ g cm$^{2}$/s.   By the end of the simulation we did not see
flattening as the clump could not cool with this
approximation.   If no angular
momentum was lost the final object would spin at 30 times the break up
speed.  Hence we expect the clump to collapse to a disk similar to the ones formed
in the isothermal run and that jets and fragmentation could occur in
nature as before.

Thus with heavy cooling suppression, beyond that expected in nature,
we retain our main result, i.e. there was still gas condensation and
the formation of clumps due to encounters.   The number and the shape
of the clumps at the end of simulation was strongly affected.  Exploring the
effect of differing thermal regulation on encounter-induced
fragmentation will be a target for future work. 

\section{Discussion and Summary}
\label{discussion}

In this work, the possibility of encounter-induced disk fragmentation
as a mechanism for forming sub-stellar objects was investigated.  The
significant increase in resolution over previous work and the use of a
realistic initial disk model ensures that the initial fragmentation to
form clumps was physically modeled within the constraints of our
thermal regulation assumptions.  These simulations have shown that 
encounters can trigger the formation of brown dwarfs as well as low
mass stars.

We found that strong instabilities can be triggered in stable disks with initial $Q_{\rm min} <
2.1$, by lowering the local Q value with large gas inflows during an 
encounter.  A relatively steep surface density profile ($\Sigma \propto
r^{-3/2}$) and a relative high ambient temperature (40 K) were used in
our simulations.  If disks have shallower density profiles, as was found in some observations of early
disks \citep{andre}, and
hence have large portions of their mass in the cooler, outer region, the
rate of fragmentation could be even higher.               

By studying collisions with random initial
orientations, it was also found that retrograde encounters (disks with retrograde spins
relative to the orbital angular momentum) more effectively transport angular
momentum, triggers gas inflows and ultimately produce
more clumps on average than the prograde ones (Figure
\ref{thetatrend}). These results differs from those reported for
simulations of star-disk encounters which indicated that prograde
encounters were more efficient in triggering disk instability
\citep[e.g.][]{boffin, lodato}.  
The difference may relate to having an extended perturber.

Dense shock layers may occur during coplanar encounters and
result in objects as seen in previous work.   However, our results indicate
that only slow coplanar encounters (cases ``Conf1'' and ``Conf2''), where the shock
layer had time to accrete substantial mass, can undergo fragmentation.
This differs from the results of 
\citet{watkinsb}, where a large fraction of their objects formed in
the fragmentation of shock layers in both coplanar and non-coplanar
cases.  This discrepancy may be due to the use of more
unstable initial disks, numerical problems related to 
lower resolution or the absence of vertical structure in the disk models
(Appendix \ref{resolution}).   However it should be noted that
coplanar encounters are very rare if the disk orientations are random.

The velocity constraint found in our simulations implies the number of
BDs produced in encounters is a strong function of the velocity
dispersion in the young star cluster.  However, the numerical trend as
a function of cluster density can be complicated.  On one hand,
more massive and denser clusters usually have larger velocity
dispersions and hence likely to produce fewer sub-stellar objects per
encounter. On the other hand, high number density will increase
the encounter rate which could potentially increase the BD numbers.

The encounter mechanism is complementary to other mechanisms for
producing BDs.  Proto-stellar disks in close proximity are naturally
produced when star cluster formation is simulated from cloud collapse.
Simulations of massive clouds typically lack the resolution necessary
to form large stable disks or to follow subsequent encounters in
detail \citep{cluster_bate}.  Thus this mechanism is compatible with
ideas such as competitive accretion and turbulent fragmentation.  The
relative importance of these processes for forming proto-stars may
influence the typical separation and orientation of proto-stellar
disks and thus affect the starting point for encounter driven object
formation.  These statistics also depend the larger environment, such
as the size, the mass and the amount of turbulent motion in the star
forming cloud.

A dynamical encounter origin for some BDs might explain observed differences in
binary statistics between stars and BDs in different cluster environments.  Some observations have
indicated that the ratio of BDs to stars is higher in denser
clusters.  For example, \citet{slesnick} found in Trapezium, the number
ratio ($R_{ss}$) of sub-stellar objects ($0.02 \lid M/M_{\sun} \lid
0.08$) over stars ($0.08 < M/M_{\sun} < 10$) is about 0.20, while
\citet{luhman03} calculated for Taurus aggregates and IC 348 $R_{ss}$
is about 0.14 and 0.12 respectively, significantly lower than
Trapezium.  The results are not conclusive as other surveys
have indicated that $R_{ss}$ in Taurus is comparable with that in
Trapezium \citep[][and references therein]{levine}.  To make further
predictions about how the BD ratio may be related to the cluster
properties, more comprehensive simulations and parameter studies would be
required, in which the encounter rate in different cluster environment
is estimated in detail and the encounter velocity parameter space is
well sampled.

Almost all the clumps formed in the isothermal simulations have
flattened shapes.  In the future evolution, we expect these clumps to
centrally condense to form a proto-BD with a disk.  Material should
accrete onto the central object on the viscous timescale.
Observations indicate that the accretion rate for low mass stars and
sub-stellar objects can be approximately fitted by $\dot{M} \sim
10^{-8} M_{\sun} \ yr^{-1} \ (M/M_{\sun})^{2}$ \citep{muzerolle}.
Hence for a relatively massive brown dwarf with 0.05 $M_{\sun}$ the
accretion rate is around $2.5 \times 10^{-11} \ M_{\sun} \
yr^{-1}$. With this rate the BD disk can have a mass as low as
$10^{-4} \ M_{\sun}$ and still have lifetime comparable to
proto-stellar disks.  To shed angular momentum, a proto-BD disk might
also launch bipolar outflows or form sub-stellar companions.  Since
most BDs formed during encounter-induced disk fragmentation have large
excess angular momenta, we expect outflows or companions to be be
common for young BD formed this way.  Recent observations have found
forbidden emission lines in BD spectra, suggesting outflows
\citep{natta}.  In particular, an outflow from a young BDs $\rho$ Oph
102 has been spatially resolved \citep{whelan}.  Some of proto-BD
disks in our simulations underwent secondary fragmentation and form
binary BD systems (cf. Section \ref{binary}).  Due to resolution and
physical limitations, this result needs to be confirmed with better
simulations.  However, the rapid rotation of these disk favours
fragmentation \citep{matsumoto} which can form companions of a range
of masses.  This provides a mechanism for forming the observed BD-BD
binaries with the small separations peaking at $\sim$ 1 to 4 AU and
with almost equal mass components \citep[][and references
therein]{whitworth}.  Planetary-mass companions have also been
observed surrounding young BD \citep[e.g. 2M1207b][]{chauvin}.

Possible future work could follow several directions.  Firstly, the
predictive power would be improved if parameters other than disk
configuration were to be well sampled, particularly the velocity of the
encounter, the disk properties and the star masses.  Combined with a
detailed calculation of encounter rates as a function of cluster
density and velocity dispersion, quantitative results could be obtained
to directly compare with the observations of BD to star ratios,
$R_{ss}$.  Secondly, when the optical-thin approximation fails in 
high density regions, the effects of shock heating and radiative
cooling become more important and should be further investigated.  Including radiative transfer
in the simulation is likely to reduce the number of objects formed per
encounter, and provide for more realistic evolution of the proto-BD
disks (as indicated in the optically thick approximation, section
\ref{cooling}).  Alternately, with simplified models, the evolution of the clump
themselves could be followed over long timescales.

\appendix 
\section{Resolution Requirements}
\label{resolution}

When designing the simulation the resolution requirement was estimated
from the \citet{bate_res} criterion, i.e., the Jeans mass should be
resolved by at least twice the number of neighbour particles,
$N_{Neighbor}$.  In our simulations we used exactly 32 neigbours.  In
the initial condition, the density of our disk varies from $10^{-13}
g/cm^{3}$ to $10^{-20} g/cm^{3}$.  Using our temperature profile,
described in section \ref{methods}, the Jeans mass at the highest
density for the initial disk was estimated to be 0.0095 $M_{\sun}$.
Hence resolving the initial disk of 0.6 $M_{\sun}$, which contains
about 63 of these smallest Jeans masses requires about 4000 particles
for each disk.  During the encounter, however, as shown in Section
\ref{results}, large inflows of gas and unstable modes increase the
density significantly.  The density typically increases by three
orders of magnitude locally where fragmentation will then occur.
Therefore to resolve the smallest Jeans mass during the simulation
significantly more particles were necessary.  To correctly simulate
gas with densities up to $10^{-10} g/cm^{3}$ about 127000 particles
were needed.  We used 200000 particles for each disk, which met the
above requirement and ensured that the fragmentation was physically
modeled.

\begin{figure*}
\includegraphics[width=\textwidth]{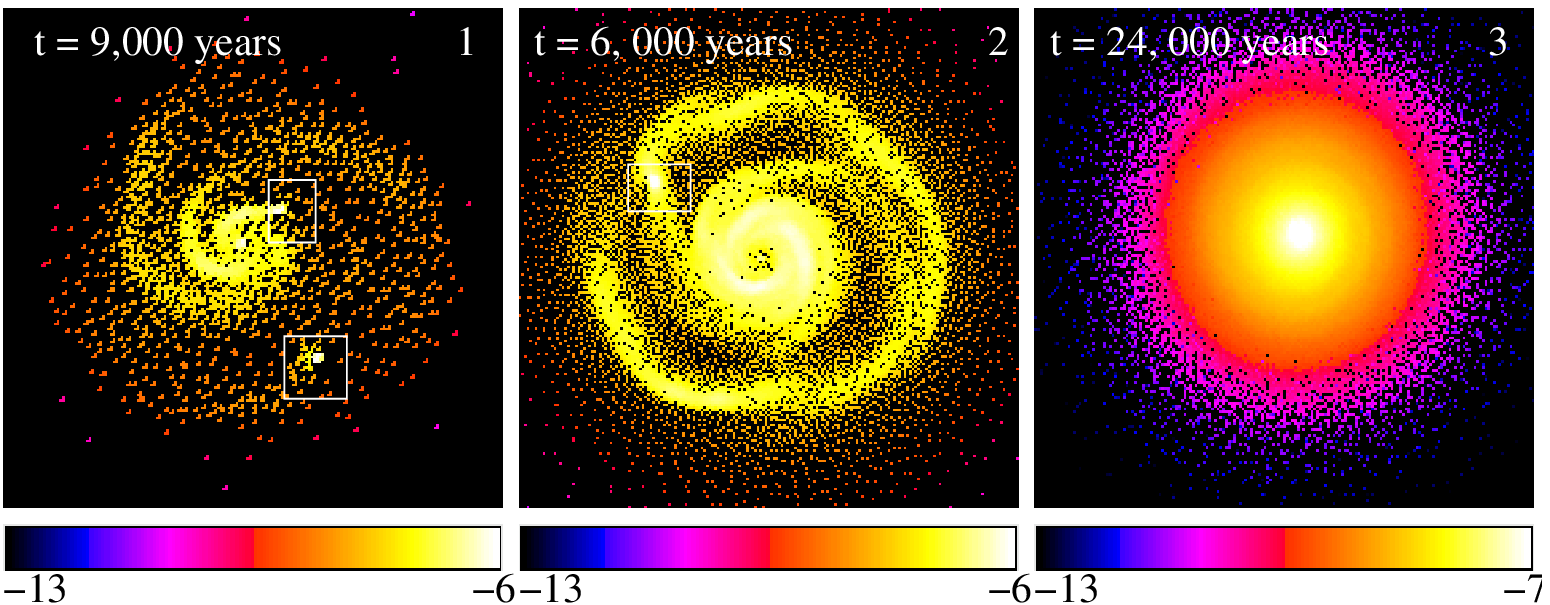}
\caption{Isolated disk simulations with different resolutions and
  disk models. All three disks have Q = 1.3 initially. Time steps for
  each snapshot are given in the upper left corners. Boxed regions
  indicate the fragments (if any). The grey scale under each panel
  gives the density ranges in units of $M_{\sun}/AU^{3}$. }
\label{res_single}
\end{figure*}

The effects of resolution and the finite scale height were studied
with two-dimensional simulations and through analytical arguments by
\cite{nelson}.  We performed two sets of comparison simulations to
investigate these effects in three dimensions.  The first set
consisted of three simulations of isolated disks with the same surface
density and temperature profile and a minimum Toomre Q of 1.3,
well above the critical value of 0.8.  The first disk was modeled by the
single layer of 2000 particles.  The second had 20000 particles but
still had no vertical structure.  The third one used 200000 particles
and the vertical structure was modeled using the methods described in
section \ref{methods}.  Figure \ref{res_single} depicts the results for
each disk at a later time step.

While no structure formed in the third, well-resolved simulation,
strong gravitational instabilities and fragmentation occurred in the
first two within several outer orbital periods.  Note that in the
first case with 2000 particles the Jeans mass was not resolved in the
high density region of the initial disk.  For the second case although
the Jeans mass was fully resolved everywhere in the initial condition
(not necessary during the simulation), the lack of vertical structure
in the model was still destabilizing and artificial fragmentation
occurred.  Although it is not clear from these two simulations which
factor (resolving Jeans mass or modeling the finite thickness) is more
important to prevent unphysical fragmentation, the test simulations
show that it is essential to consider both factors in the simulation.
    
\begin{figure*}
\includegraphics[width=\textwidth]{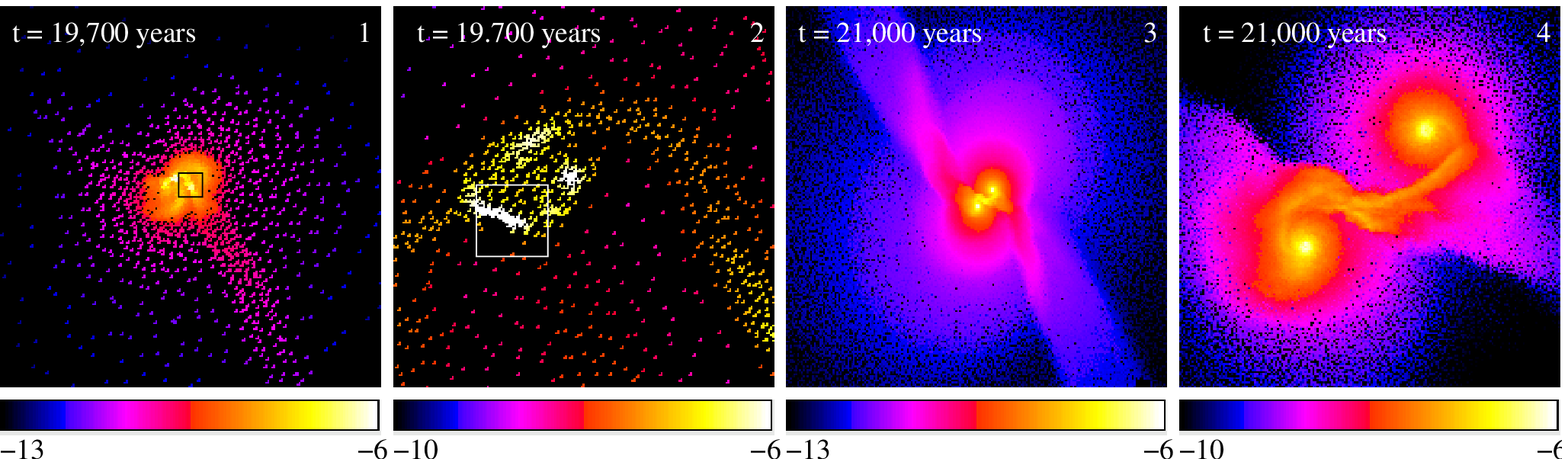}
\caption{Disk encounter simulations with different resolutions and
  disk models. The disks have Q = 2.1 initially. Time steps for
  each snapshot are given in the upper left corners. The first two
  panels (panel 1-2) depicted the low resolution run and panel 3-4
  are from the higher resolution one. Boxed regions
  indicate the fragments (if any). The grey scale under each panel
  gives the density ranges in units of $M_{\sun}/AU^{3}$. }
\label{res_encounter}
\end{figure*}

The second series consist of two simulations of retrograde dynamical
encounters of disks with Q = 2.1. The disks were very stable in
isolated runs. The dynamical parameters of the encounter was the same
as case ``Q3'' in Table \ref{list}. In the first simulation the
disks were modeled by single layer of 2000 particles, while in the second
the disk was resolved by 200000 particles and had vertical
structures. The results is depicted in Figure \ref{res_encounter}. 

Similar to the isolated runs, without properly resolving the Jeans
mass and scale height, the encounter triggered fragmentation at
time step of 19700 years in the shock layer between two coplanar disks
(Panel 2, Figure \ref{res_encounter}). In the second simulation,
although the shock layer was still produced (Panel 4), no
fragmentation occurred in the layer. The encounter was also modeled
using 400000 particles with the same result.
       
In summary, regardless of whether the disks are passively evolving or undergo
a physical impact, numerical effects may be significant if the Jeans mass
is not resolved or if no vertical structure is presented in the disk
model. The major numerical effect is then enhanced instability
and the formation of artificial clumps.  Our resolution of 200000
particles was chosen based on a conservative estimate of
resolution requirements, e.g., it assumed the density of the whole disk could
increase up to $10^{-10} g/cm^{3}$.  In fact,
most outer region of the disk had density below $10^{-12} g/cm^{3}$ when
fragmentation initial occurred.  With this resolution it was ensured
that the initial disk fragmentation (if any) was physical.  We also
performed simulations at higher resolution (400000 and 800000
particles) for some specific cases and the results converged.

\end{document}